\documentclass[twocolumn]{aastex62}
\usepackage[toc,page]{appendix}

\newcommand{\tr}{t_{\rm cool}/t_{\rm ff}}

\definecolor{webgreen}{rgb}{0,.5,0}
\definecolor{webbrown}{rgb}{.6,0,0}

\graphicspath{{./}{figures/}}

\begin{document}

\title[Environmental Dependence of Feedback]{Environmental Dependence of Self-Regulating Black-hole Feedback in Massive Galaxies}

\correspondingauthor{Deovrat Prasad}
\email{deovratd@msu.edu}

\author[0000-0003-1255-6375]{Deovrat Prasad}
\affiliation{Department of Physics and Astronomy, Michigan State University, MI, US}

\author[0000-0002-3514-0383]{G. Mark Voit}
\affiliation{Department of Physics and Astronomy, Michigan State University, MI, US}

\author[0000-0002-2786-0348]{Brian W. O'Shea}
\affiliation{Department of Physics and Astronomy, Michigan State University, MI, US}

\affiliation{Department of Computational Mathematics, Science, and Engineering and National Superconducting Cyclotron Laboratory, Michigan State University, MI, US}

\author[0000-0002-6837-8195]{Forrest Glines}
\affiliation{Department of Physics and Astronomy, Michigan State University, MI, US}

\affiliation{Department of Computational Mathematics, Science, and Engineering, Michigan State University, MI, US}

\begin{abstract}
In the universe's most massive galaxies, active galactic nuclei (AGN) feedback appears to limit star formation. The accumulation of cold gas near the central black hole fuels powerful AGN outbursts, keeping the ambient medium in a state marginally unstable to condensation and formation of cold gas clouds. However, the ability of that mechanism to self-regulate may depend on numerous environmental factors, including the depth of the potential well and the pressure of the surrounding circumgalactic medium (CGM). Here we present a suite of numerical simulations, with halo mass ranging from $2 \times 10^{12} \, M_\odot$ to $8 \times 10^{14} \, M_\odot$, exploring the dependence of AGN feedback on those environmental factors. We include the spatially extended mass and energy input from the massive galaxy's old stellar population capable of sweeping gas out of the galaxy if the confining CGM pressure is sufficiently low. Our simulations show that this feedback mechanism is tightly self-regulating in a massive galaxy with a deep central potential and low CGM pressure, permitting only small amounts of multiphase gas to accumulate and allowing no star formation. In a similar mass galaxy with shallower central potential and greater CGM pressure the feedback mechanism is more episodic, producing extended multiphase gas and allowing small rates of star formation ($\sim 0.1 \, M_\odot \, {\rm yr}^{-1}$). At the low-mass end, the mechanism becomes implausibly explosive, perhaps because the CGM initially has no angular momentum, which would have reduced the amount of condensed gas capable of fueling feedback.    
\end{abstract}


\section{Introduction}
\label{sec:introduction}

X-ray observations during the past two decades have revolutionized the astronomical community's understanding of how AGNs regulate cooling, condensation, and star formation at the centers of galaxy clusters. The high X-ray surface brightness of a cool-core cluster allows detection of cavities created in the hot gas by AGN outflows, along with measurements of their size, which show that AGN kinetic power is comparable to the radiative cooling rate of the cluster core (\citealt{Churazov2001, McNamara2007}).  Among galaxy clusters in the nearby universe, high-power radio AGNs, multiphase gas, and star formation are found only in those with low-entropy gas at their centers (\citealt{cavagnolo09, Cavagnolo2008, rafferty2006, Sun2009, Hoffer2012, Rawle2012}). These relationships indicate that AGN heating self-regulates through a feedback loop in which accretion of cold clouds condensing out of the hot medium strongly boosts AGN feedback power, thereby maintaining the core in a state marginally unstable to precipitation of cold clouds.  The minimum ratio of gas cooling time ($t_{\rm cool}$) to free-fall time ($t_{\rm ff}$) in such a self-regulating system tends to be in the range $10 < \min(\tr) < 20$ (\citealt{mccourt12, sharma12, gaspari2012, Prasad15, Prasad2018, li15,  voit15N, voit15L, voit2017}).  Many cool-core clusters observed with {\em Chandra} do indeed fall into this range (\citealt{voit15N, Hogan2017}), strongly supporting the hypothesis that AGN feedback in cluster cores is throttled by a transition to precipitation.

These findings raise an important question: Does precipitation-regulated AGN feedback also limit cooling and condensation of circumgalactic gas around smaller galaxies?  The hot atmospheres of those galaxies tend to be harder to observe with X-ray telescopes, but preliminary investigations indicate that the AGN feedback loop observed in the universe’s largest galaxies may also limit the density of circumgalactic gas all the way down through Milky-Way scales (\citealt{Voit2018_LX-T-R,Voit2019_pNFW}). X-ray observations of massive ellipticals \citep{werner2012, werner2014} show that the lower bound on $t_{\rm cool}$ in the ambient medium tracks the $\min(\tr) \sim 10$ locus across two orders of magnitude in radius, from $\sim$20 kpc down to $\sim$200 pc \citep{voit15N, voit2015, voit2020}. Also, the general features of precipitation-regulated star formation are consistent with several of the major scaling relations observed among galaxies (\citealt{voit15L}).

Several different channels can feed accretion of cold gas onto the central black hole.  Numerical simulations show that cold gas can stream along cosmological dark matter filaments and accrete onto a halo's central galaxy if those cold streams are not disrupted by a surrounding hot gaseous halo (\citealt{keres2005,dekel2009}).  In a galaxy cluster, radiative cooling of the halo's hot gas can produce a central cooling flow (\citealt{fabian1994, pizz05}).  And in massive elliptical galaxies, accumulations of gas ejected from the old stellar population can also produce a central cooling flow \citep{mathewandloewenstein1986,vd2011}.  All of these supply channels are capable of fueling star formation at rates greatly exceeding those observed in massive elliptical galaxies, requiring a feedback mechanism to quench star formation. 

Numerical simulations demonstrate that bipolar AGN jets fuelled by cold accretion can effectively quench star formation (\citealt{gaspari2012,gaspari2013,li2014,li15,Prasad15,Prasad2018,yang2016,meece2017}).  
However, both the amount of cold gas that accumulates and its spatial distribution appear to depend on galactic environment.  In galaxy clusters, $10^9$--$10^{11} \, M_\odot$ of molecular gas can accumulate, extending to tens of kiloparsecs from the center.  Less molecular gas accumulates at the centers of smaller halos, and its spatial distribution appears to depend on the galaxy's stellar velocity dispersion.  In massive elliptical galaxies with central velocity dispersion $\sigma_v < 240$ km s$^{-1}$, the cold gas typically extends beyond the central 2~kpc, but it tends to be more centrally concentrated in galaxies with $\sigma_v > 240$ km s$^{-1}$, as long as they are not at the centers of galaxy clusters \citep{voit2015,voit2020}.  This relationship between $\sigma_v$ and the distribution of cold gas is intriguing, particularly in light of optical observations showing that nquenching of star formation is more closely related to the stellar velocity dispersion of the central galaxy than to any other observable property \citep{Wake2012,Bluck2016,Terrazas2016,Bluck2020}.

Here, we present a suite of 3-D hydrodynamic simulations motivated by the observed relationships between AGN feedback and galactic environment.  The simulations were designed to investigate how bipolar AGN outflows fuelled by cold accretion depend on halo mass, central velocity dispersion, and the pressure of the circumgalactic medium (CGM). Section \ref{sec:valve} provides additional background on how the central stellar velocity dispersion is thought to influence the relationship between AGN feedback and CGM pressure. Section \ref{sec:NM} describes the numerical setup for our simulations. Section \ref{sec:init} specifies the initial conditions for each numerical experiment. Section \ref{sec:results} presents the main results.  Section \ref{sec:limit} discusses the limitations of our simulations and their potential effects on the results.  With those limitations in mind, \S \ref{sec:disc} discusses our results, comparing them to theoretical models and prior simulations. Section \ref{sec:conc} summarizes the paper's main findings.

\section{The Black-Hole Feedback Valve}
\label{sec:valve}

Spatially extended gas and energy input from the central galaxy's old stellar population is an essential feature of the simulation suite we are presenting, because of how it links CGM pressure to AGN feedback.  In a massive elliptical galaxy, energy input from SNIa heating cannot by itself push ejected stellar gas out of the galaxy's halo.  Instead, that gas accumulates until its ambient density becomes great enough for radiative cooling to exceed supernova heating \citep{mathewandloewenstein1986,Ciotti1991,binneyandtabbor1995}. That process can take a few billion years. But eventually, the resulting cooling flow should make AGN feedback the dominant heating mechanism. It is therefore rather surprising that X-ray observations of nearby elliptical galaxies with $\sigma_v > 240 \, {\rm km \, s^{-1}}$ are often consistent with models of steady subsonic outflows driven by SNIa heating, at least within the central 1--10~kpc \citep{voit2015, voit2020}.

In those models, AGN feedback and SNIa heating play complementary roles. Feedback from the AGN is necessary to lift circumgalactic gas out of the galaxy's potential well. However, lifting of the CGM lowers its pressure, causing conditions within the galaxy to change.  As AGN feedback drives CGM pressure down, gas pressure and density within the galaxy also decline. 
AGN feedback therefore enables SNIa heating to become more competitive with radiative cooling within the galaxy.  Once the galaxy's ambient gas density becomes low enough for SNIa heating to exceed radiative cooling, SNIa heating can limit the cooling flow that powers AGN feedback.  Stellar mass and energy sources can therefore couple with AGN feedback to maintain a nearly steady state in which SNIa heating slightly exceeds radiative cooling of gas within the galaxy. 

\citet{voit2020} refers to this tuning mechanism as a ``black-hole feedback valve."  It operates when the specific energy of gas ejected from the old stellar population ($\epsilon_*$) is not much greater than the square of the galaxy's stellar velocity dispersion ($\sigma_v$).  In a steady outflow driven by SNIa heating, the gradients of gas pressure, density, and entropy depend on the ratio $\epsilon_*/\sigma_v^2$.  As that ratio declines, the proportion of SNIa energy needed to lift gas out of the galaxy becomes greater and the gradients of gas properties become larger.  

A simple calculation in \citet{voit2020} demonstrates that the tuning mechanism for a galaxy with a stellar population age $\sim 10$~Gyr works best if $\sigma_v > 240 \, {\rm km \, s^{-1}}$.  At that limiting value of $\sigma_v$, the gradient of specific entropy\footnote{This paper quantifies specific entropy in terms of the entropy index $K \equiv kTn_e^{-2/3}$.} approximately corresponds to $K \propto r^{2/3}$, making the pressure and density gradients approximately $\propto 1/r$.  For greater values of $\sigma_v$, the density gradient is steeper, causing radiative cooling to exceed SNIa heating at small radii, even if SNIa heating exceeds radiative cooling at larger radii.  The result is a cooling flow at small radii ($< 1$ kpc) surrounded by a slow SNIa-heated outflow extending to beyond $\sim 10$~kpc.  Consequently, the CGM pressure confining the slow outflow determines the cooling-flow rate at small radii.  That coupling is what enables the mechanism to tune itself.  It should inevitably shut down star formation by suppressing accumulation of extended multiphase gas in galaxies with a central velocity dispersion exceeding the critical limiting value, as long as the AGN produces enough power to lift the CGM out of the halo's potential well.

One of this paper's motivations was to see whether this black-hole feedback valve mechanism would naturally arise in a 3-D numerical simulation with the required properties.  That is why $\sigma_v$ is an important environmental parameter and also why we pay particular attention to the relationship between SNIa heating and radiative cooling. Our attempt to replicate the mechanism was only partially successful. Section \ref{sec:limit} discusses some future improvements to simulations that may help to replicate the mechanism with greater realism.\\

\section{Numerical Setup}
\label{sec:NM}
We modified \textsc{Enzo}, an adaptive mesh refinement (AMR) code \citep{Bryan2014,Enzo_2019}, to simulate AGN and stellar feedback in idealized galactic environments across a broad mass range.    The masses of the simulated halos span the range $2\times10^{12} M_\odot \leq M_{200} \leq 8\times10^{14} M_{\odot}$, where $M_{200}$ is the mass contained within the radius $r_{200}$ encompassing a mean mass density 200 times the cosmological critical density.
We solve the standard hydrodynamic equations in Cartesian coordinates, including radiative cooling, gravity, star formation, stellar feedback, and AGN feedback, along with mass and energy input from an old stellar population (see \S \ref{sec:Afeed} for details). These simulations employ the piecewise parabolic method (PPM) with a Harten-Lax-van Leer-Contact (HLLC) Riemann solver. 

\subsection{Grids}
\label{sec:bc}
The simulation domain is a $(4\ {\rm Mpc})^3$ box with a $64^3$ root grid and up to 10 levels of refinement. 
The central (128 kpc)$^3$ region enforces static regions of grid refinement ranging from level 6 to level 9, with the refinement level increasing toward the center.  In the central $(2 \, {\rm kpc})^3$, the mesh is fixed to be at the highest level of refinement.  
This design ensures that the CGM is highly resolved at all times with a minimum cell size $\Delta l \approx 61 \, {\rm pc}$ and a maximum cell size $\Delta l < 1$ kpc.

\subsection{Environmental Parameters}
\label{sec:gpt}
The gravitational potentials in our simulations have three components, and they do not change with time.  We use an NFW form (\citealt{nav1997}) for the dark-matter halo, with mass density $\rho_{\rm DM} \propto r^{-1} (1 + r/r_{\rm s})^{-2}$, where $r_{\rm s}$ is the NFW scale radius and $c_{200} = r_{200} / r_{\rm s}$ is a halo concentration parameter.  For the central galaxy, we use a Hernquist profile (\citealt{hern1990}), with a stellar mass density $\rho_* \propto r^{-1} (1 + r/r_{\rm H})^{-3}$, where $r_{\rm H}$ is the Hernquist scale radius.  The potential of the central supermassive black hole of mass $M_{\rm BH}$ follows a Paczynski-Witta form (\citealt{pacwita1980}).  
\begin{table*}
\begin{center}
\caption{Galactic Environmental Parameters}
\begin{tabular}{ c  c  c  c  c  c  c  c  c }
\hline
Galaxy Model &  $M_{200}$  & $c_{200}$ & $r_{\rm 200}$ 
     & $M_*$ & $r_{\rm H}$ & $\sigma_{v}$ & $M_{\rm BH}$ & Analog Galaxy \\
     & ($M_{\odot}$) &          &   (kpc)   
     & ($M_{\odot}$) &  (kpc)   & (km s$^{-1}$) & $(10^8 M_\odot)$ &               \\
\hline
BGC (Brightest Cluster Galaxy) 
    & $8\times10^{14}$ & 4.9 & 1920 & 3$\times10^{12}$ & 10 & 230 & 50 & ...  \\
MPG (Multiphase Galaxy)
    & $4.4\times10^{13}$ & 9.5 & 730 & 1.2$\times10^{11}$ & 1.2  & 230 & 4.6 & NGC 5044  \\
SPG (Single Phase Galaxy)
    & $4.0\times10^{13}$ & 7.5 & 700 & 2$\times10^{11}$ & 1.6  & 280 & 26 & NGC 4472  \\
SEG (Smaller Elliptical Galaxy)
    & $2\times10^{12}$ & 5 & 275 & 1$\times10^{11}$ & 1.5  & 150 & 0.7 & ...  \\
\hline
\end{tabular}
\label{tab:gal_params}
\end{center}
\end{table*}

Table~\ref{tab:gal_params} provides the parameter values for each model.  Two of those models are based on particular galaxies.  The multiphase galaxy (MPG) model is intended to resemble NGC~5044, which has a central stellar velocity dispersion\footnote{Central stellar velocity dispersions quoted in this paper are from Hyperleda: http://leda.univ-lyon1.fr/.} $\sigma_v \approx 225 \, {\rm km \, s^{-1}}$, and represents the population of massive elliptical galaxies with extended multiphase gas.  The single phase galaxy (SPG) model is intended to resemble NGC~4472, which has a central stellar velocity dispersion $\sigma_v \approx 282 \, {\rm km \, s^{-1}}$, and represents the population of massive elliptical galaxies without much multiphase gas beyond the central kiloparsec.  The brightest cluster galaxy (BCG) model is meant to represent the central region of a typical massive galaxy cluster.  The smaller elliptical galaxy (SEG) model is designed to test how the AGN feedback mechanism used in the larger halos operates when the halo mass and central stellar velocity dispersion are reduced.

\subsection{Star Formation}
\label{sec:SF}

A simulation cell forms a star particle if its gas satisfies several criteria based on the \citet{cenos92} prescription:
\begin{itemize}
\item The baryon density must exceed a threshold density ($\sim 1 \, {\rm cm^{-3}}$). 
\item The flow must be converging ($\nabla \cdot \vec{v_b} < 0$).
\item The gas must be cold ($T < 1.1\times10^4$ K).
\item The gas mass of the cell ($m_{\rm b}$) must exceed $10^3 \, M_\odot$. 
\item The cooling time of the gas must be less than the dynamical time for that cell's gas, $t_{\rm dyn} = \sqrt{3\pi/(32 G \rho_{\rm gas})}$. 
\end{itemize} 
A star particle is then formed with mass $m_* = f_{*,{\rm eff}}  ( {\Delta t} /{t_{\rm dyn}} ) m_{\rm b}$, where  the star-formation efficiency parameter is set to $f_{*,{\rm eff}} = 0.1 $ in this simulation suite.  

\subsection{Stellar Mass and Energy Input}
\label{sec:Afeed}

The central galaxy's stars heat the gas through two separate channels:
\begin{enumerate}
\item New stars forming during the course of the simulation produce supernovae (SNII) that impart both thermal energy and momentum.
\item Old stars with the density distribution of the Hernquist potential (\S \ref{sec:gpt}) add heat through SNIa explosions and thermalization of stellar kinetic energy.
\end{enumerate}

Feedback from stars formed during the simulation follows the prescription from \citet{Bryan2014}. After a star particle of mass $m_*$ forms, a fraction $f_{m,*}$ of its mass is added back to the cell, along with thermal energy $E_{\rm SN} = f_{\rm SN} m_* c^2$. Our simulations adopt the parameter values $f_{\rm SN}=1\times10^{-5}$ and $f_{m,*}= 0.25$. The returned gas formally has a metallicity $f_{Z,*}=0.02$, which can be used as a passive tracer but is not included in our radiative cooling calculations.  This process starts immediately after the formation of the star particle and decays exponentially, with a time constant of 1~Myr. 

SNIa heating is modelled with steady, spherically symmetric injection of thermal energy into the simulation domain at a rate proportional to the stellar mass density. The total energy ejected from SNIa explosions assumes $10^{51}$ erg of per SNIa at a specific rate of $3 \times 10^{-14} \, {\rm SNIa \, yr^{-1} \, M_\odot^{-1}}$, following \citealt{voit2015}.  At this rate, an old stellar population of mass $\sim 10^{11} \, M_\odot$ adds $\sim 10^{41} \, {\rm erg \, s^{-1}}$ of thermal energy.
The old stellar population also injects kinetic energy as it sheds gas mass in the form of stellar winds and SNIa explosions.  Our simulations assume that this kinetic energy immediately thermalizes.  We assume a specific gas ejection rate $\alpha_* = 10^{-19} \, {\rm s}^{-1}$, such that the net ejected matter per unit time per unit volume is $\alpha_* \rho_*$.  To simplify the calculation of thermalized kinetic energy, we assume an isotropic 1-D stellar velocity dispersion of $300$~km~s$^{-1}$ at all radii in all of our runs, following \citealt{wang2019}.  The difference between this uniform value of $\sigma_v$ and the actual one is inconsequential, because energy input from SNIa heating is several times ($\gtrsim 5$) greater than the kinetic energy injection in all cases.

\subsection{AGN Feedback}
\label{sec:agn}
AGN feedback is introduced into the simulation using a feedback zone attached to the AGN particle (\citealt{meece2017}), which is always located at the geometric center of the halo. We drive AGN feedback in the form of a bipolar outflow by putting source terms for mass, momentum and energy into the fluid equations as follows:
\begin{eqnarray}
\label{eq:fl_eq}
\frac {\partial \rho} { \partial t} + \nabla \cdot (\rho {\bf u}) & \, = \, & S_{\rho} \nonumber \\
\frac {\partial (\rho {\bf v})} { \partial t} + \nabla \cdot (\rho {\bf v} \otimes {\bf v}) & \, = \, & -\nabla P - \rho\nabla \Phi + S_{\bf p} \nonumber \\
\frac {\partial (\rho e)} {\partial t} + \nabla \cdot [(\rho e +P){\bf v}] &  \, = \, & \rho{\bf v \cdot g} - n_e n_i \Lambda(T,Z) + S_{e} \nonumber
\end{eqnarray}
where $S_{\rho}$, $S_{\bf p}$ and $S_{e}$ are the density, momentum and energy source terms, respectively.  The specific energy $e$ includes kinetic energy, and the corresponding equation of state is $P = (\gamma -1) \rho (e - v^2/2)$.

\subsubsection{Accretion and AGN Feedback Efficiency}
\label{sec:acc}
The accretion rate $\dot{M}_{\rm acc}$ onto the central supermassive black hole is calculated by assuming that all the cold gas ($T<10^5$ K) within $r< 0.5$ kpc accretes onto the central black hole on a 1~Myr time scale.  
This mass accretion rate fuels AGN feedback at an energy output rate given by:
\begin{equation}
\dot{E}_{\rm AGN} = \epsilon_{\rm AGN} \dot{M}_{\rm acc} c^2
\; \; ,
\end{equation}
where $c$ is the speed of light and the feedback efficiency parameter $\epsilon_{\rm AGN}$ is taken to be $10^{-4}$ for all our runs with AGN feedback. A cold gas mass equal to  $\dot{M}_{\rm acc} \Delta t$ is removed from the spherical accretion zone ($r < 0.5$ kpc) by subtraction of gas mass from each cell in that zone with a temperature below $10^5$~K, and the amount of gas mass subtracted from the cell is proportional to its total gas mass.

\subsubsection{Feedback Energy Deposition}
\label{sec:part}

The AGN output energy is partitioned into kinetic and thermal parts and introduced using the source terms described above.  The source regions are cylinders of radius 0.5~kpc extending along the jet axis from $r = 0.5$~kpc to $r = 1$~kpc in each direction. Each jet therefore subtends 1~radian at $r = 1$ kpc.

In each cell of volume $\Delta V_{\rm cell}$ within the source region of volume $V_{\rm jet}$, the density source term is $S_\rho = \Delta m_{\rm cell} / \Delta V_{\rm cell}$, where $\Delta m_{\rm cell} = (\Delta V_{\rm cell} / V_{\rm jet}) \dot{M}_{\rm acc} \Delta t$ and $\Delta t$ is the time step.  
Within the source region, the corresponding energy source term in each cell is $S_e = (\Delta e_{\rm cell} + \Delta {\rm KE}_{\rm cell})/V_{\rm cell}$, where 
\begin{equation}
   \Delta e_{\rm cell} = (1 - f_{\rm kin}) \dot{E}_{\rm AGN} \Delta t \cdot \frac{\Delta V_{\rm cell}}{V_{\rm jet}} 
\end{equation}
\begin{equation}
   \Delta {\rm KE}_{\rm cell} = f_{\rm kin} \dot{E}_{\rm AGN} \Delta t \cdot \frac{\Delta V_{\rm cell}}{V_{\rm jet}}
   \; \; .
\end{equation}
For all our runs the ratio of kinetic to total AGN output energy is fixed at $f_{\rm kin} = 0.9$.

The momentum source term in each cell is $S_{\bf p}= \Delta {\bf p}_{\rm cell}/ \Delta V_{\rm cell}$, where
\begin{equation}
    \Delta {\bf p}_{\rm cell} = \Delta m_{\rm cell} 
        \sqrt{ \frac{2\Delta {\rm KE}}{{\Delta m_{\rm cell}}} } \hat{\bf n}
\end{equation}
and $\hat{\bf n}$ is a unit vector pointing away from the origin along the jet injection axis. 

\section{Atmospheric Initial Conditions}
\label{sec:init}

\begin{table*}
\begin{center}
\caption{Feedback Parameters, Atmospheric Initial Conditions, and Simulation Outcomes}
\begin{tabular}{c c c c c c c c c c c c}
\hline
Galaxy Model & Feedback Model & $\epsilon_{\rm AGN}$ 
    & $K_0(t=0)$ & $K_{100}(t=0)$ & $\alpha$ & run time & State at $>1$ Gyr \\
             &                &        
    &  (${\rm keV \, cm^2}$) & (${\rm keV \, cm^2}$) &         &   (Gyr)  &                     \\
\hline
BCG & AGN + Stellar & $10^{-4}$ & 15  & 230 & 1.1  & 2 & episodically regulated core  \\
MPG & Stellar       &  0        & 1.3 & 150 & 1.05 & 1.5 & cooling flow   \\
MPG & AGN + Stellar & $10^{-4}$ & 1.3 & 150 & 1.05 & 1.5 & episodically regulated core  \\
SPG & Stellar       &  0        & 1.5 & 400 & 1.05 & 2 & cooling $>$ heating  \\
SPG & AGN + Stellar & $10^{-4}$ & 1.5 & 400 & 1.05 & 1.5 & steadily regulated core  \\
SEG & Stellar       &  0        & 5   &  85 & 1.1  & 1.5 & cooling flow       \\
SEG & AGN + Stellar & $10^{-4}$ & 5   &  85 & 1.1  & 1   & over-heated core   \\
\hline
\end{tabular}
\end{center}
\label{tab:initcond}
\end{table*}

\begin{figure*}
\centering
    \includegraphics[width=0.47\textwidth]{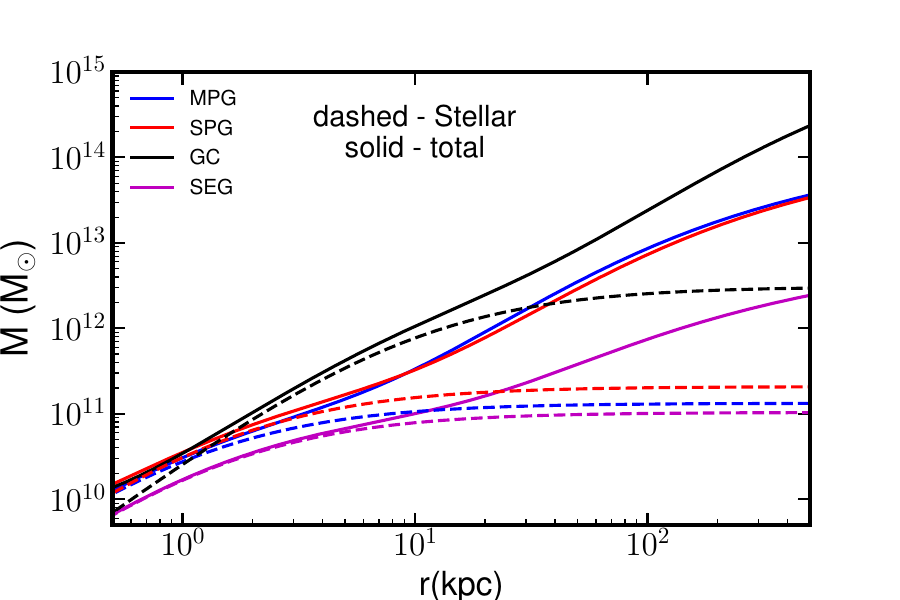}
    \includegraphics[width=0.47\textwidth]{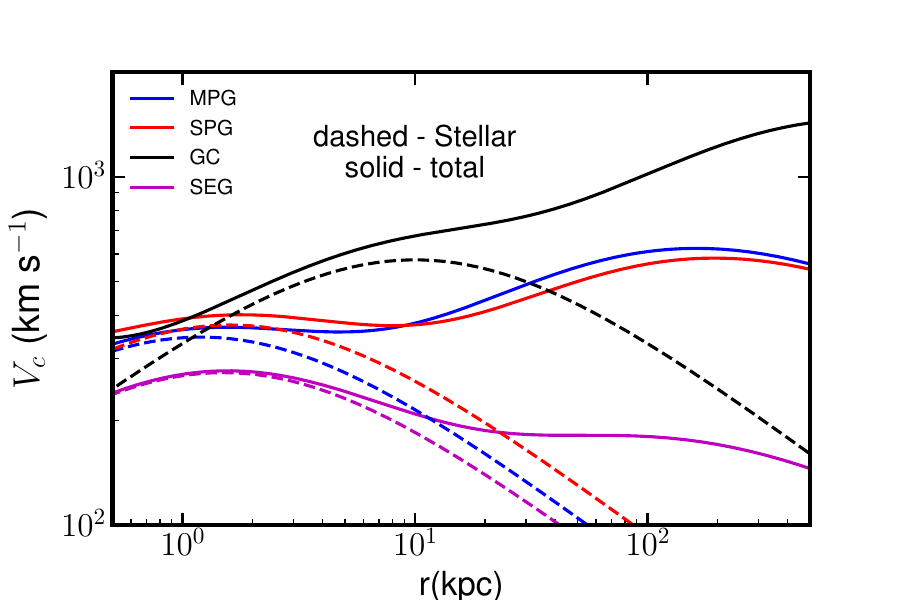}
    \includegraphics[width=0.47\textwidth]{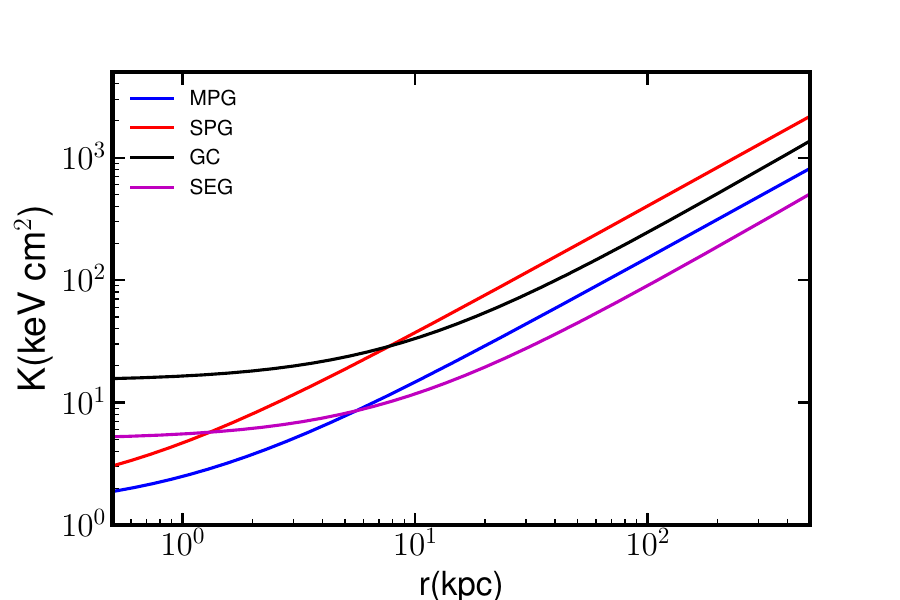}
    \includegraphics[width=0.47\textwidth]{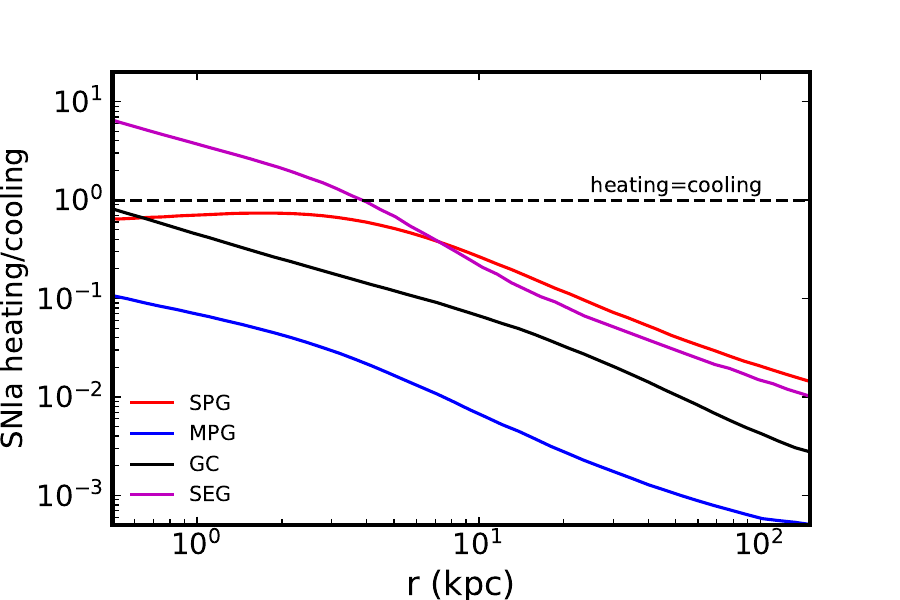}
    \caption{Environmental and initial atmospheric conditions.  Top panels show mass enclosed within radius $r$ (left) and the corresponding circular velocity (right), with solid lines showing total mass and dashed lines showing just the stellar component.  
    Bottom panels show each atmosphere's initial entropy profile (left) and ratio of SNIa heating to radiative cooling (right).  Each color represents a different galaxy model as shown in the legend, except for the dashed black line in the lower-right panel, which shows where SNIa heating equals cooling.
    }
    \label{fig:initial}
\end{figure*}

All of the simulations in this paper begin with hydrostatic, single-phase galactic atmospheres having profiles of density, pressure, and specific entropy consistent with observations of nearby counterparts.  Figure~\ref{fig:initial} shows the initial atmospheric properties for each simulation, along with the properties of the gravitational potentials and central galaxies outlined in Table \ref{tab:gal_params}.  Solid lines in the upper panels of Figure \ref{fig:initial} show the total mass $M(r)$ enclosed within radius $r$ and the corresponding circular velocity, $v_c = \sqrt{G M(r)/r}$, while dashed lines show the same quantities for just the stellar mass component.  Each initial entropy profile is modelled using the form $K(r)=K_0 + K_{100} (r/100 \, {\rm kpc})^\alpha$ (\citealt{cavagnolo09}).  Table~\ref{tab:initcond} gives the starting values of $K_0$, $K_{100}$ and $\alpha$ for each halo, which are based on observations by \citet{Cavagnolo2009} for the BCG model, \citet{werner2012,werner2014} for the MPG and SPG models, and \citet{Lakhchaura2018,babyk2018} for the SEG model.

The bottom left panel shows each galaxy's initial entropy profile, and the bottom left panel shows the initial ratio of SNIa heating to radiative cooling at each radius.  Initial atmospheric pressure in the BCG and MPG is large enough that radiative cooling significantly exceeds SNIa heating everywhere.  In the initial state of the SPG, SNIa heating nearly equals radiative cooling inside of $\sim 5$~kpc but is less significant at larger radii.  However, the lower-pressure atmosphere of the smaller elliptical galaxy (SEG) allows SNIa heating to exceed radiative cooling within the central $\sim 5$~kpc.
In all cases, the initial electron number density is set to a constant value ($n_e =5\times 10^{-6} \, {\rm cm^{-3}}$) at an outer domain radius corresponding to $2 r_{200}$ in the three lower-mass halos and to $r_{200}$ in the most massive halo.  

Throughout the simulation, atmospheric gas is allowed to cool to $10^3$ K using tabulated \cite{SD1993} cooling functions with one-third solar metallicity for the BCG simulation and solar metallicity for all other runs.

\section{Results}
\label{sec:results}

This section describes the key results from our simulations. We first examine how the atmospheres evolve without AGN feedback. In each case a cooling flow results, even if SNIa heating initially exceeds radiative cooling.  Then we analyze how each atmosphere changes when AGN feedback is active.  We find that the three more massive systems each settle into a self-regulated state within $\sim 1$~Gyr.  The fluctuations in AGN feedback are larger in the two massive halos with greater CGM pressure and lower central velocity dispersion ($\sigma_v \approx 230 \,{\rm km \, s^{-1}}$), causing larger changes in core conditions and producing more multiphase gas over a larger region.  In contrast, the massive halo with lower CGM pressure and greater central velocity dispersion ($\sigma_v \approx 280 \,{\rm km \, s^{-1}}$) quickly settles into a nearly steady self-regulating state in which SNIa heating exceeds radiative cooling within the central $\sim 10$~kpc.  However, the lowest-mass halo ($M_{200} = 2 \times 10^{12} M_\odot$) fails to self-regulate because AGN feedback becomes too explosive.

\subsection{Simulations without AGN feedback}
\label{sec:nojet}

\begin{figure}
    \includegraphics[width=0.48\textwidth]{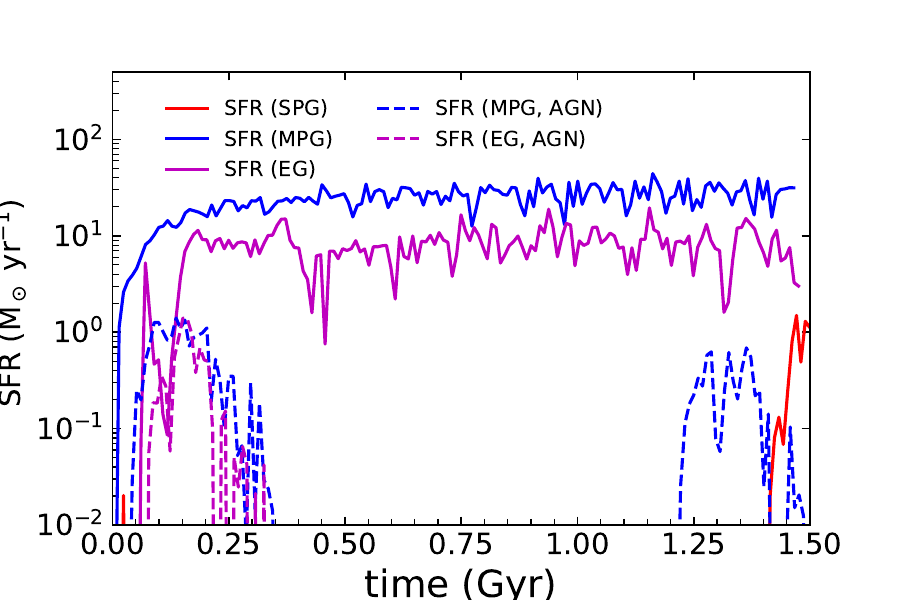}
    \includegraphics[width=0.48\textwidth]{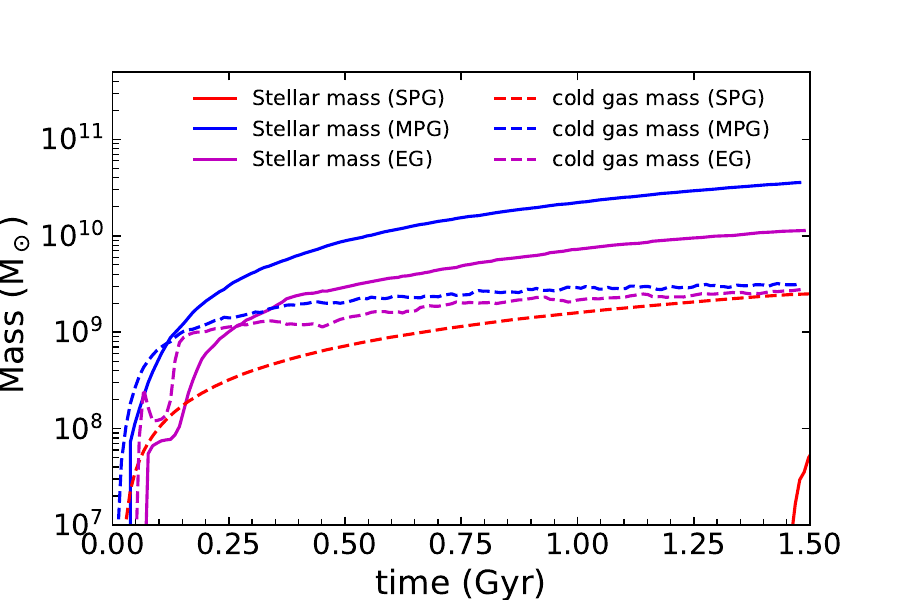}
    \caption{Star formation rates (SFR, top panel) and accumulations of cold gas and new stars (bottom panel) in the SPG (red), MPG (blue), and SEG (purple) models. Star formation rates are smoothed on a 10~Myr timescale for runs without AGN feedback (solid lines) and with AGN feedback (dashed lines), and star formation begins promptly in both the MPG and SEG models. It does not begin in the SPG model without AGN feedback until $t \approx 1.4$~Gyr but then rises to a steady rate $\sim 5 \, M_\odot \, {\rm yr}^{-1}$ by $t \approx 1.75$ Gyr.  The bottom panel shows accumulated mass in stars (solid lines) and cold gas (dashed lines) as functions of time for only the models without AGN feedback. 
    }
    \label{fig:nj}
\end{figure}

To see how quickly star formation begins and what its rate would be without AGN feedback, we ran simulations of the three smaller halos with $\epsilon_{\rm AGN} = 0$. We did not perform a similar simulation of the BCG, because stellar feedback is obviously insufficient to limit star formation in such a massive halo. Figure \ref{fig:nj} shows the resulting star formation rates as functions of time.  Unsurprisingly, the MPG model begins to form stars almost immediately, because radiative cooling exceeds SNIa heating at all radii.  More than $10^9 \, M_\odot$ of cold gas ($T<10^5$ K) accumulates by $t \approx 250$~Myr, and star formation then proceeds at a steady rate $\sim 25 \, M_\odot \, {\rm yr}^{-1}$.  According to \citet{voit2011}, the steady cooling-flow rate associated with an entropy profile $K(r) \propto r$ is
\begin{eqnarray}
    \dot{M} & \; = \; & \frac {8 \pi} {3} \mu m_p (kT)^2 \Lambda(T) 
                    \left( \frac {K} {r} \right)^{-3} \\
            & \; \approx \; & 
                 24 \, M_\odot \, {\rm yr}^{-1} 
                    \left( \frac {kT} {1 \, {\rm keV}} \right)^2
            \left[ \frac {\Lambda (T)} {10^{-23} \, {\rm erg \, cm^3 \, s^{-1}}} \right]
            \label{eq:steadyCF}
            \\ \nonumber
            & &  \; \; \; \; \times
            \left( \frac {K/r} {1 \, {\rm keV \, cm^2 \, kpc^{-1}}} \right)^{-3}
\end{eqnarray}
in an isothermal potential.  The asymptotic star formation rate observed in the MPG simulation without AGN feedback is therefore consistent with its initial entropy profile, which is $K/r \approx 1 \, {\rm keV \, cm^2 \, kpc^{-1}}$ at $\sim 10$~kpc.

More surprisingly, star formation also begins promptly in the SEG simulation without AGN feedback, even though SNIa heating initially exceeds radiative cooling out to $\sim 5$~kiloparsecs from the center.  Figure \ref{fig:nj} shows that star formation rises to $\sim 8 \, M_\odot \, {\rm yr^{-1}}$ within the first 200~Myr. The amount of cold gas that accumulates during that same time period is comparable to the steady-state amount in the MPG simulation and is similar to the amount of hot gas that starts the simulation with a cooling time $t_{\rm cool} \lesssim 200$ Myr.   
The asymptotic star-formation rate is again consistent with equation (\ref{eq:steadyCF}), but with $kT \sim 0.3$~keV.

Cooling and star formation begins within $\sim 100$~Myr in the SEG model despite the central SNIa heating, because the weight of its CGM prevents the gas ejected by old stars from leaving the galaxy. The initial gas mass density at $\sim 1$~kpc is $\rho \approx 5 \times 10^{-26} \, {\rm g \, cm^{-3}}$.  At similar radii, the stellar mass density is $\rho_* \approx 10^{-22} \, {\rm g \, cm^{-3}}$, meaning that stellar ejecta can double the gas-mass density there on a timescale $\rho (\alpha_* \rho_* )^{-1} \sim 150$~Myr if the gas cannot be pushed outward. Some of the gas is pushed outward, but not enough to prevent a buildup of gas within the central 5~kpc.  Meanwhile, radiative cooling of gas just beyond $\sim 5$~kpc produces an entropy inversion that makes the atmosphere convectively unstable and promotes thermal instability.  Condensing gas clouds then sink to the center and initiate star formation.  The resulting SNII explosions briefly suppress additional star formation but result in uplift of low-entropy ambient gas that precipitates at $\sim 10$~kpc, forming new cold clouds.  Those clouds then rain down toward the galaxy's center, and during the next 50~Myr the ambient gas settles into a steady cooling flow.   

Star formation requires more time to reach a steady state in the SPG simulation without AGN feedback. A brief burst of star formation happens at $t \sim 50$~Myr, because radiative cooling initially exceeds SNIa heating everywhere.  SNII feedback from that initial burst then lowers the central density, which allows SNIa heating to exceed radiative cooling out to $\sim 1$~kpc. Star formation remains suppressed for the next $\sim 1.3$~Gyr, while the central gas pressure gradually rises. The central pressure goes up during this period because SNIa heating cannot push ejected stellar gas outward as fast as it accumulates and also because cooling of the overlying layers increases their weight.  Eventually, the central gas density becomes great enough for radiative cooling to exceed SNIa heating, and the resulting cooling flow boosts the star formation to a steady state rate of $\sim 3.5 $ M$_\odot$ yr$^{-1}$ at $t \sim 2.0 \, {\rm Gyr}$.

\begin{figure*}
    \includegraphics[width=2.4in,height=1.8in]{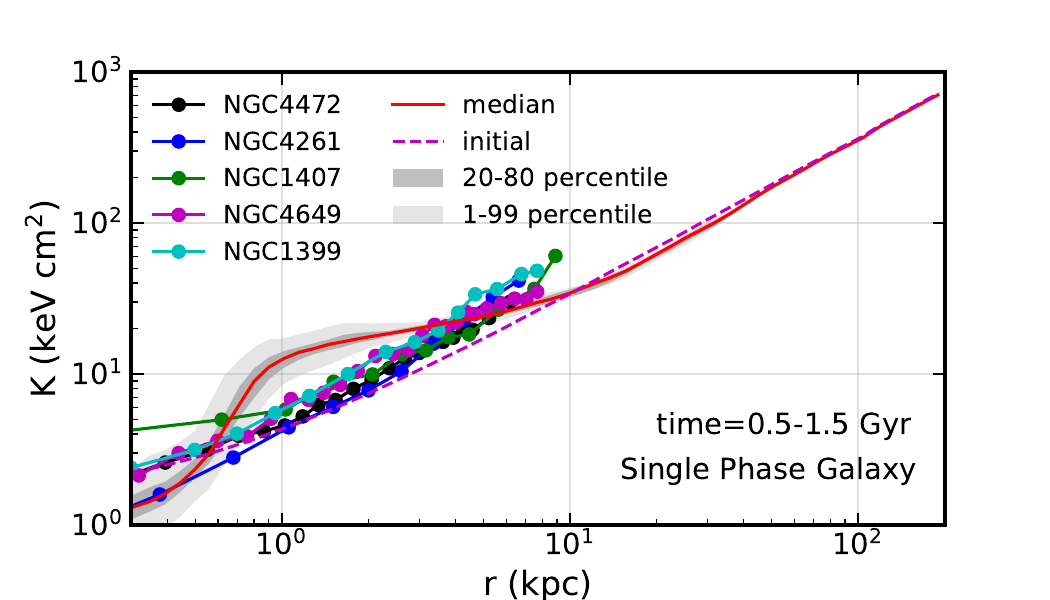}
    \includegraphics[width=2.4in,height=1.8in]{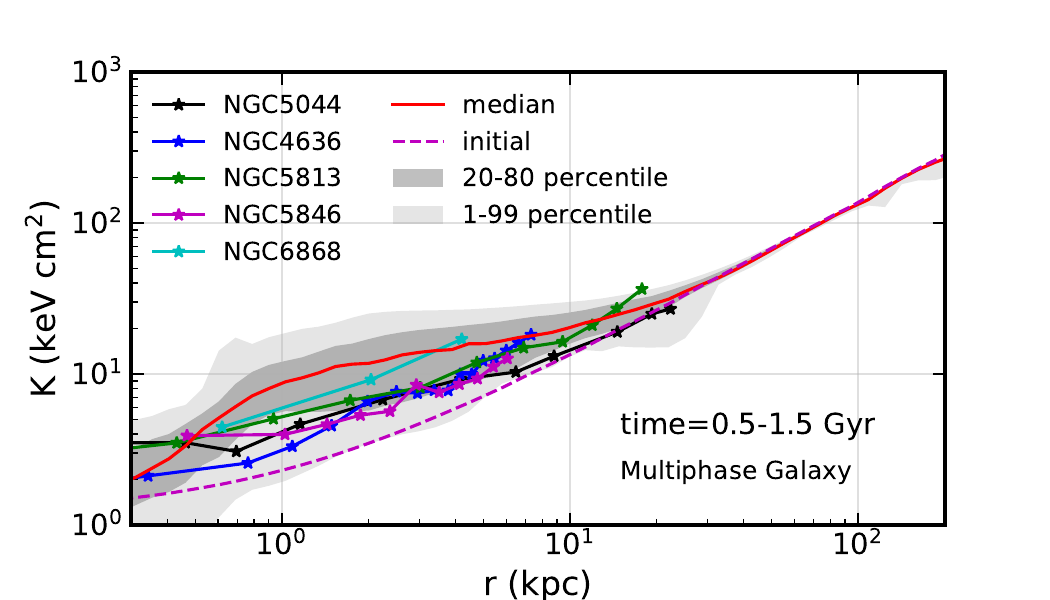}
    \includegraphics[width=2.4in,height=1.8in]{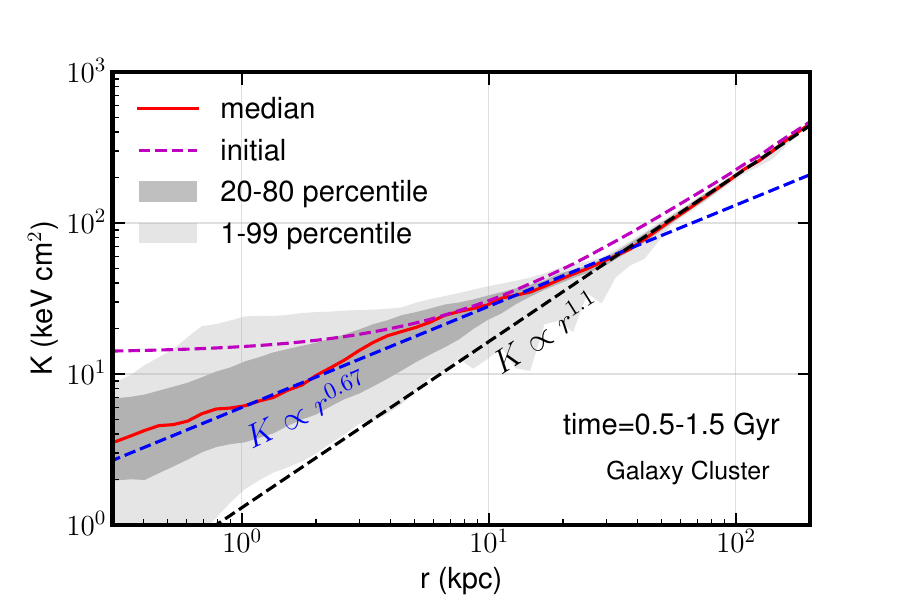}
    \caption{Emissivity-weighted median entropy profiles for the single phase galaxy (SPG, left panel), multiphase galaxy (MPG, center panel), and brightest cluster galaxy (BCG, right panel), simulated with AGN feedback. In each panel, a solid red line shows the median entropy profile during the period $t = 0.5$--1.5~Gyr, and a dashed magenta line shows the initial profile. Dark grey shading shows the 20th to 80th percentile range of the median entropy profile during that period.  Light grey shading shows the 1st to 99th percentile. Symbols connected by lines of other colors show observed entropy profiles of representative single phase (left panel) and multiphase (center panel) galaxies from (\citealt{werner2012, werner2014}).  
    Dashed lines in the right panel indicate the approximate slope of the inner part ($K \propto r^{0.67}$, blue) and outer part ($K \propto r^{1.1}$, black) of the BCG's median entropy profile.
    }
    \label{fig:entropy}
\end{figure*}

\subsection{Massive halos with AGN feedback}

In all of our simulations with AGN feedback, star formation is highly suppressed relative to the respective no-AGN counterparts (see Figure~\ref{fig:nj}).  However, condensation of the ambient medium couples with AGN feedback differently, depending on both the depth of the central potential well the and atmospheric pressure at larger radii. In the SPG simulation, coupling between condensation and AGN feedback is remarkably tight and maintains a nearly steady feedback-regulated state.  In contrast, the MPG and BCG simulations exhibit greater feedback bursts.  This section examines how these three massive halos self-regulate, while \S \ref{sec:small_halo} looks at what happens in the SEG simulation, which fails to self-regulate. 

\subsubsection{Radial profiles}
\label{sec:rad_prof}

Figure \ref{fig:entropy} shows the median emissivity-weighted radial entropy profiles in the SPG, MPG, and BCG simulations with AGN feedback.  A dashed magenta line indicates the initial entropy profile in each simulation.  A solid red line traces the median profile for the entire period from 0.5 to 1.5~Gyr.  Dark grey shading shows the 20--80 percentile range of the median entropy profile during that period, and light grey shading shows the 1--99 percentile range. 

In each case, the median entropy profile shifts from its initial state into a different self-regulated state.  The BCG entropy profile settles from a flat-entropy core into a self-regulated state with $K \propto r^{2/3}$ in the central $\sim 30$~kpc.  The MPG entropy profile rises to a self-regulated state with a mean entropy at $< 10$~kpc several times greater than the initial state.  The SPG entropy profile also rises within the central $\sim 10$~kpc by a factor of $\sim 2$. However, the shading shows that tightly-coupled feedback confines the SPG entropy profile to a much narrower range than in the BCG or MPG.

Comparing the data in Figure~\ref{fig:entropy} with the simulation results reveals a significant discrepancy in the vicinity of 1~kpc, where specific entropy in the SPG and MPG simulations exceeds the observations by a factor of 2 to 3.  In the middle panel showing the MPG simulation, some of that discrepancy might arise from a selection effect.  The galaxies shown have particularly bright X-ray emission produced by atmospheres denser than average for their mass, meaning that they have lower than average specific entropy, but the data still remain within the fluctuation range of the simulation.  However, the data in the SPG panel are well outside the fluctuation range of the SPG simulation.  We therefore suspect that the entropy excess near 1~kpc results from a limitation of our simulation, which is the thickness of the AGN jet there.  Section~\ref{sec:limit} discusses that limitation in more detail.

\begin{figure*}
    \includegraphics[width=1.12\textwidth]{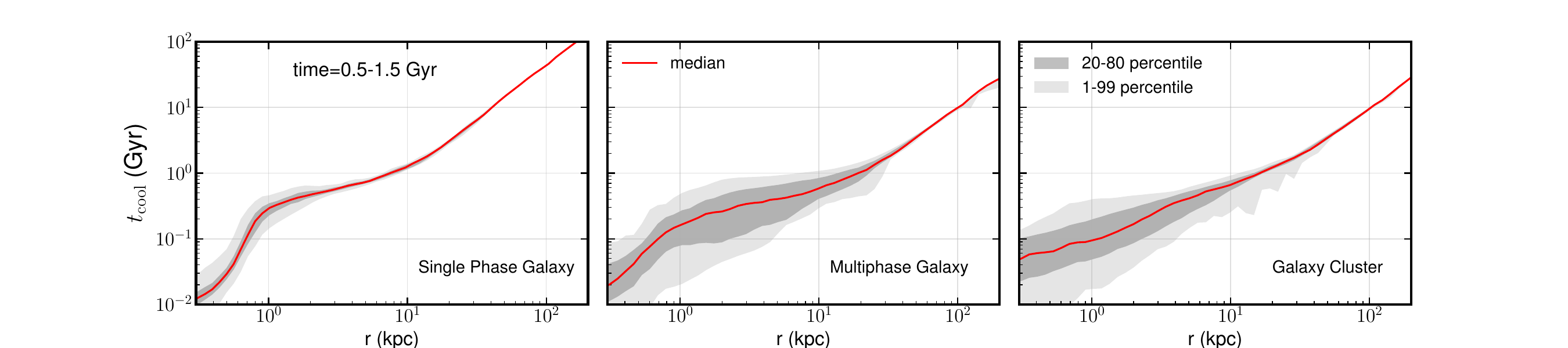}
    \includegraphics[width=1.12\textwidth]{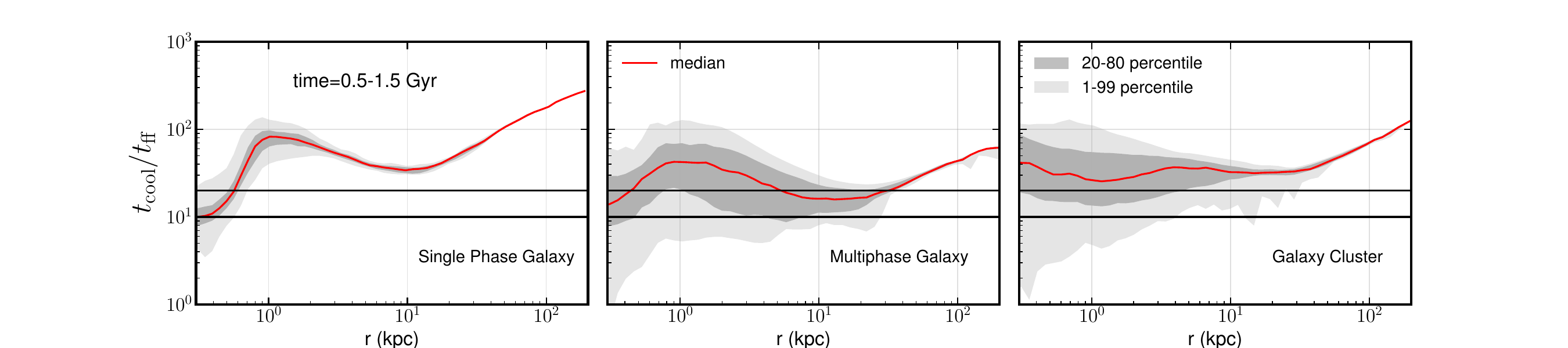}
    \caption{ Median radial profiles of $t_{\rm cool}$ (upper panels) and $t_{\rm cool} / t_{\rm ff}$ (lower panels) for simulations with AGN feedback.
    Red lines and grey shading have the same meanings as in Figure~\ref{fig:entropy}.  Black lines in the lower panels indicate the range $10 < t_{\rm cool} / t_{\rm ff} < 20$ characteristic of precipitation-regulated feedback.  The single phase galaxy (left panels) remains above that range outside of 1~kpc, with $\min (t_{\rm cool} / t_{\rm ff}) \sim 30$.  The multiphase galaxy (middle panels) self-regulates with $10 < t_{\rm cool} / t_{\rm ff} < 20$ near 10~kpc during most of the 0.5--1.5~Gyr period, with brief excursions down to $\min (t_{\rm cool} / t_{\rm ff}) \sim 6$ at smaller radii.  The brightest cluster galaxy fluctuates in and out of the $10 < t_{\rm cool} / t_{\rm ff} < 20$ range, with brief excursions to $\min (t_{\rm cool} / t_{\rm ff}) \sim 5$ within the central few kiloparsecs.
    }
    \label{fig:tcool}
\end{figure*}

Figure \ref{fig:tcool} shows the median radial profiles of both $t_{\rm cool}$ (upper panels) and the $t_{\rm cool}/t_{\rm ff}$ ratio (lower panels) during the same time period for the same three simulations.  The left panels show that the SPG simulation self-regulates with $\min( t_{\rm cool} / t_{\rm ff} ) > 20$ at $> 1$~kpc and $t_{\rm cool} > 1$~Gyr at 10~kpc.  The middle panels show that the MPG self-regulates with most of its time spent in the $10 < \min(t_{\rm cool} / t_{\rm ff}) < 20$ range, with $t_{\rm cool} < 1$~Gyr at 10~kpc.  The right panels show that the BCG self regulates with $\min(t_{\rm cool} / t_{\rm ff})$ fluctuating in and out of that range, also with $t_{\rm cool} < 1$~Gyr at 10~kpc.

\subsubsection{Self-regulation and SNIa heating}
\label{sec:temp_evol}

\begin{figure*}
    \includegraphics[width=2.4in,height=1.6in]{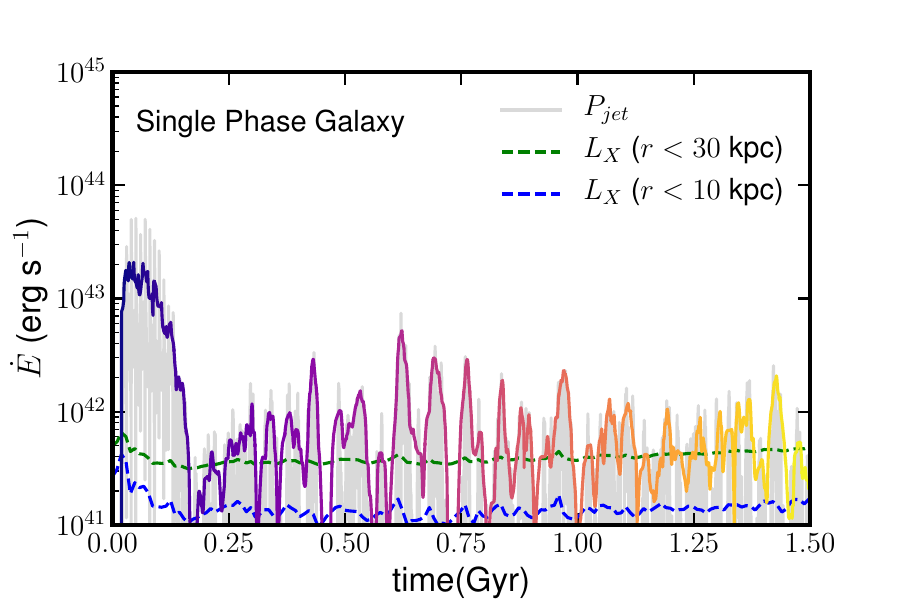}
    \includegraphics[width=2.4in,height=1.6in]{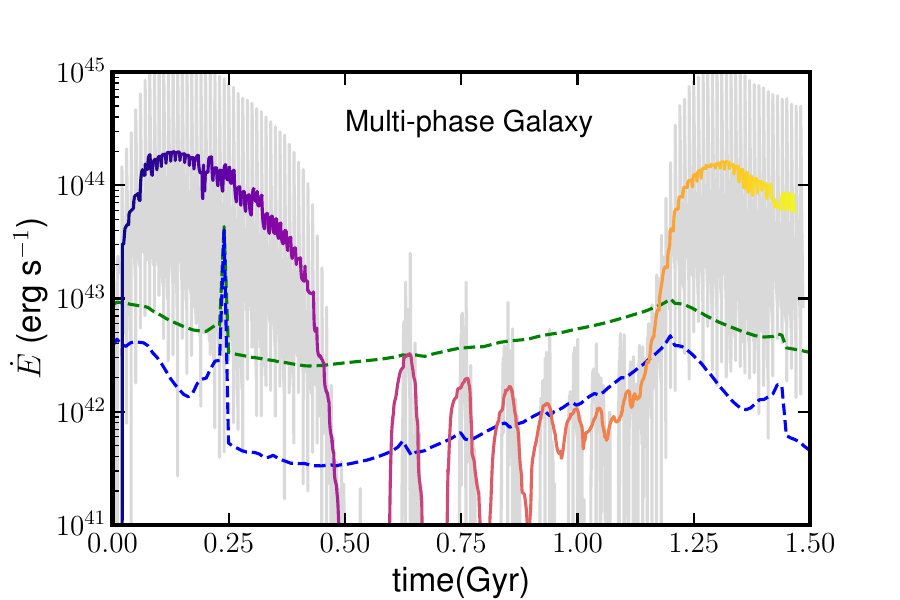}
    \includegraphics[width=2.7in,height=1.6in]{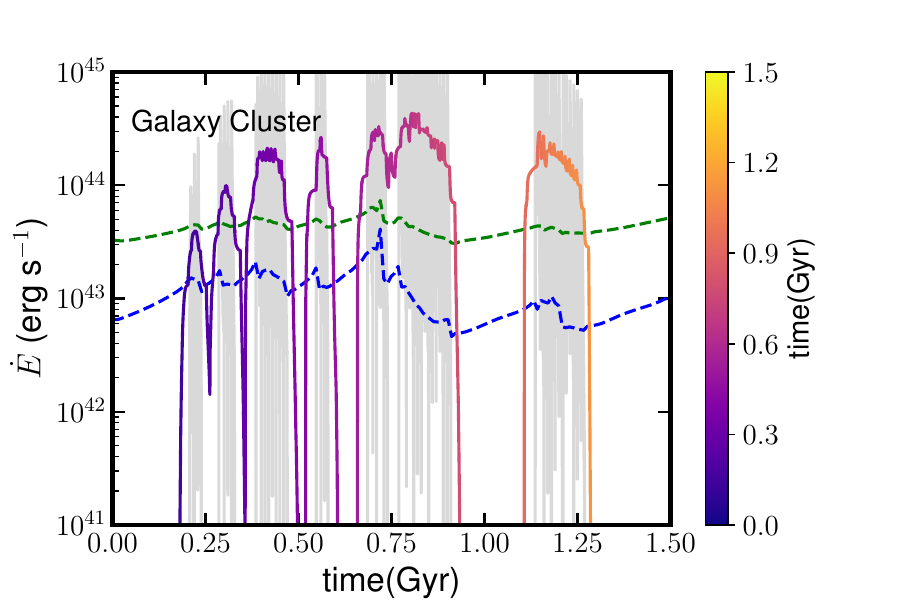}    
    \includegraphics[width=2.4in,height=1.6in]{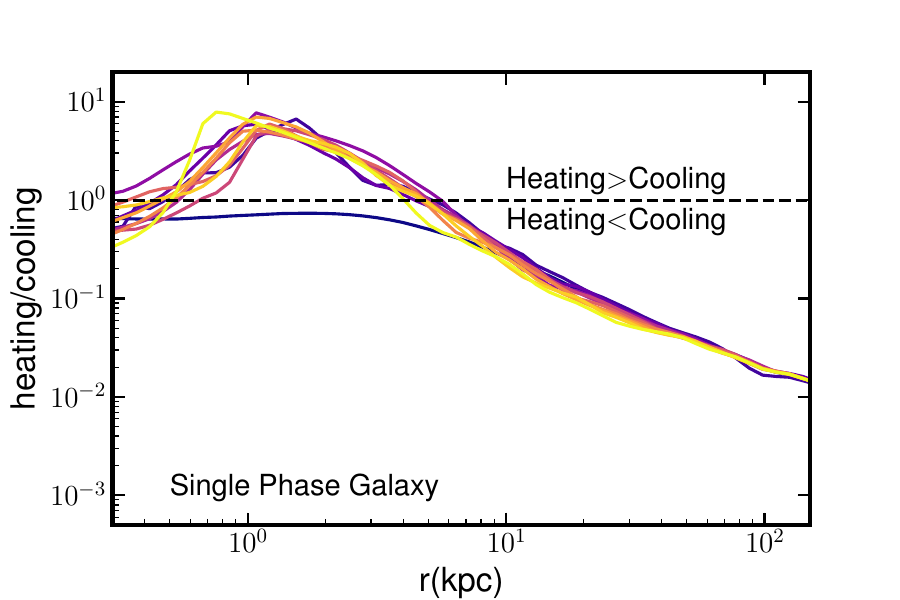}
    \includegraphics[width=2.4in,height=1.6in]{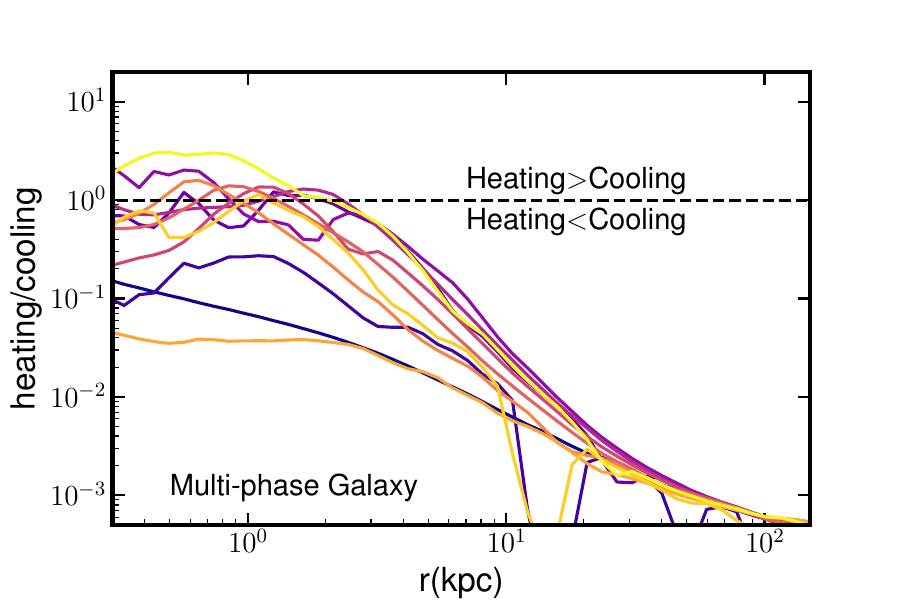}
    \includegraphics[width=2.7in,height=1.6in]{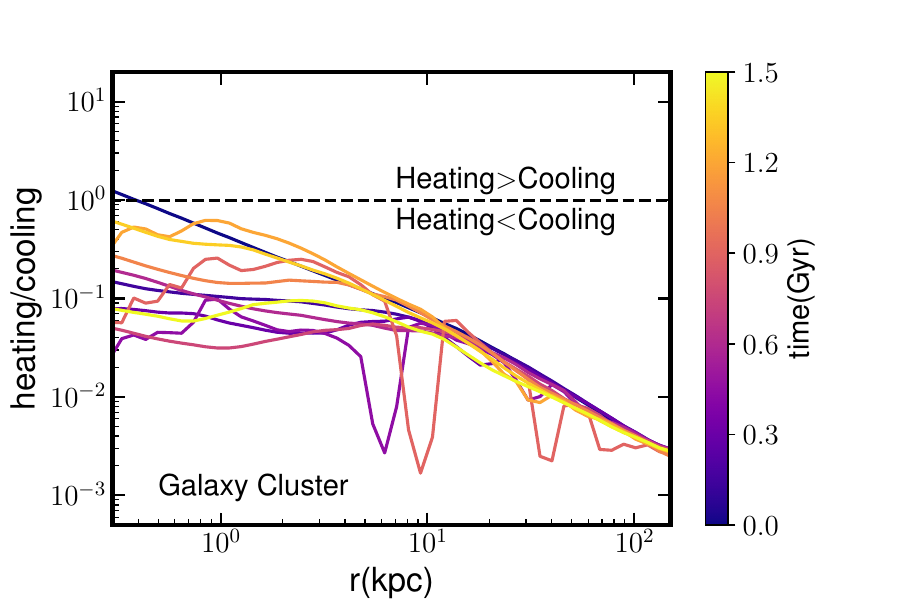}
    \caption{{\it Upper Panels}: Jet power ($P_{\rm jet}$) and X-ray luminosity ($L_X$; 0.5-7 keV) as functions of time for simulations of the single phase galaxy (left), multiphase galaxy (center) and brightest cluster galaxy (right) with AGN feedback.  Grey lines show instantaneous $P_{\rm jet}$ and solid lines with changing color with time show $P_{\rm jet}$ smoothed on a $20$ Myr timescale. Dashed lines show radiative losses from the inner 10~kpc (blue) and inner 30~kpc (green). All three simulations self-regulate but exhibit two different feedback modes: a high-power mode ($P_{\rm jet} \sim 10^{43}$--$10^{44} \, {\rm erg \, s^{-1}}$) capable of altering the central 30~kpc, and a low-power mode ($P_{\rm jet} \sim 10^{41}$--$10^{42} \, {\rm erg \, s^{-1}}$) that cannot compensate for cooling in the halos of the MPG or BCG.
    {\it Lower panels}: Ratio of SNIa Heating (SN Ia) to radiative cooling as a function of radius every 150 Myr during the evolution of each halo. In the single phase galaxy (left), AGN feedback promptly lowers the atmosphere's density, enabling SNIa heating to exceed radiative cooling from $\sim 0.5$~kpc to $\sim 5$~kpc.  That state corresponds in time to the steady low-power mode of self-regulation. The high-power feedback mode in the multiphase galaxy (center) expands the galactic atmosphere, lowering its X-ray luminosity until SNIa heating becomes comparable to radiative cooling near $\sim 1$~kpc.  AGN feedback then switches to a low-power mode that is insufficient to replace radiative losses within $\sim 30$~kpc, causing feedback to revert to a high-power mode at 1.2~Gyr, when radiative cooling once again exceeds SNIa heating everywhere. In the BCG simulation (right), radiative cooling rapidly exceeds SNIa heating everywhere, fueling only the high-power feedback mode.
    }
    \label{fig:lx_pjet}
\end{figure*}

Self-regulation in these three simulations depends on how multiphase condensation couples AGN feedback with the state of the ambient medium.  The upper panels in Figure \ref{fig:lx_pjet} show how injected jet power ($P_{\rm jet}$) is related to X-ray luminosity  ($L_X$) over the time period 0--1.5~Gyr.The X-ray luminosity is calculated in the `yt' package (\citealt{YT} ) using the `Cloudy' emission table for temperatures in range $0.5-7$ keV with solar metallicity for all runs except the galaxy cluster case, where a one-third solar metallicity emission table is used. After the initial AGN outburst of $\sim 10^{43} \, {\rm erg \, s^{-1}}$, the single phase galaxy settles into a nearly steady state with time-averaged AGN jet power that is similar to the X-ray luminosity from within the central 30~kpc.  Frequent bursts fuelled by fluctuations in the amount of cold gas within the central 0.5~kpc cause jet power to vary by a factor $\sim 10$, but the power output remains steady when smoothed over timescales $> 100$~Myr.  However, the other two simulations experience much greater fluctuations in jet power.

In the multiphase galaxy, AGN feedback is bimodal. The simulation starts with an outburst of jet power $\sim 10^{44}$ erg s$^{-1}$. Heat input from that outburst causes the galaxy's atmosphere to expand, lowering its density and significantly reducing radiative cooling of the central 30~kpc. The AGN then enters a low-power state with $P_{\rm jet}$ fluctuating on a $\sim 100$~Myr timescale between $< 10^{41} \, {\rm erg \, s^{-1}}$ and a few times $10^{42} \, {\rm erg \, s^{-1}}$.  Meanwhile, the atmosphere's X-ray luminosity climbs, because time-averaged AGN power is much less than $L_X$ from within 30~kpc.  Those radiative losses allow the weight of the CGM to compress the galactic atmosphere, gradually raising its density and pressure. The AGN remains in this low-power mode for $\approx 800$~Myr but then reverts back to a high-power state, similar to the initial one, for another $\approx 200$~Myr.

The BCG simulation remains in a state similar to the high-power mode of the MPG simulation most of the time and does not have a low-power mode.  It is either near $P_{\rm jet} \sim 10^{44} \, {\rm erg \, s^{-1}}$ or at $P_{\rm jet} < 10^{41} \, {\rm erg \, s^{-1}}$.  The state of extremely low power is likely to be artificial, resulting from the fact that our feedback algorithm sets AGN power to zero if there is no cold gas within the central 0.5~kpc.  A more realistic model would include AGN power resulting from Bondi-like accretion of hot ambient gas, but jet power in that mode would be far too low to significantly affect the surrounding atmosphere. 

The lower panels of Figure \ref{fig:lx_pjet} show how the mode of AGN feedback is related to the ratio of SNIa heating to radiative cooling in the central few kiloparsecs. The radiative cooling for each radial shell is calculated $\sim n_e n_i \Lambda(T) dV$, where $dV$ is the volume of each radial shell and $\Lambda(T)$ is the tabulated Sutherland-Dopita cooling function (\citealt{SD1993}). Initially, radiative cooling in the single phase galaxy is slightly greater than SNIa heating within $\sim 5$~kpc of the center.  Cooling of that gas fuels a $\sim 100$~Myr burst of feedback that lowers the central gas density until SNIa heating exceeds radiative cooling from $\sim 0.5$~kpc to $\sim 5$~kpc.  AGN feedback then enters the low-power mode, fueled only by cooling of gas within $\sim 0.5$~kpc of the center.  And that mode is sufficient to keep the single-phase galaxy in a steady state, for at least $\sim 1.5$~Gyr.

In the multiphase galaxy, a larger initial burst of AGN power is needed to lower the atmosphere's density because its confining CGM pressure is greater.  However, SNIa heating becomes comparable to radiative cooling at $\sim 1$~kpc by $t \approx 300$~Myr.  The simulation then settles into the low-power feedback mode for nearly 1~Gyr. During that time, AGN power is less steady than in the single phase galaxy because near-equality of SNIa heating and radiative cooling at $r < 3$~kpc allows larger condensation events to intermittently feed the AGN.

The BCG simulation, on the other hand, is in a cooling-dominated state everywhere during virtually the entire time period.  AGN feedback cannot lower the atmosphere's central density enough for SNIa heating to equal radiative cooling.  Therefore, the mode of self-regulation connects AGN feedback to large condensation events, which occur well outside of the central kiloparsec.

\subsubsection{Cold Gas and Star Formation}
\label{fig:gas_and_stars}

\begin{figure*}
    \includegraphics[width=0.33\textwidth]{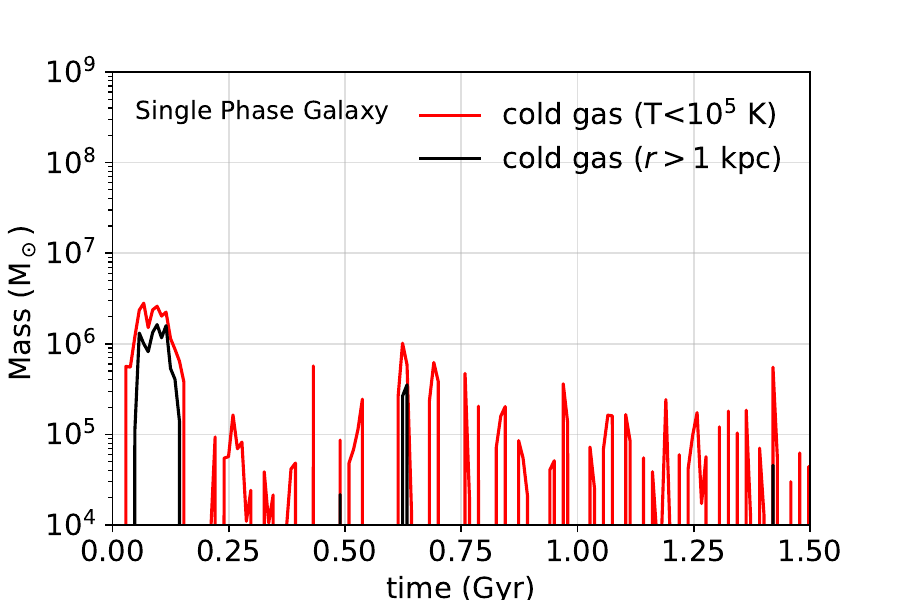}
    \includegraphics[width=0.33\textwidth]{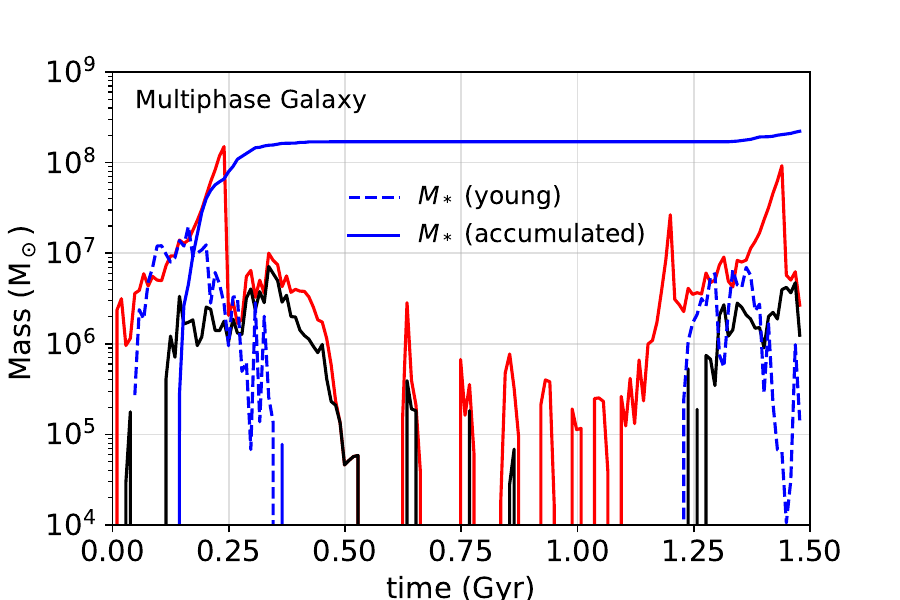}
    \includegraphics[width=0.33\textwidth]{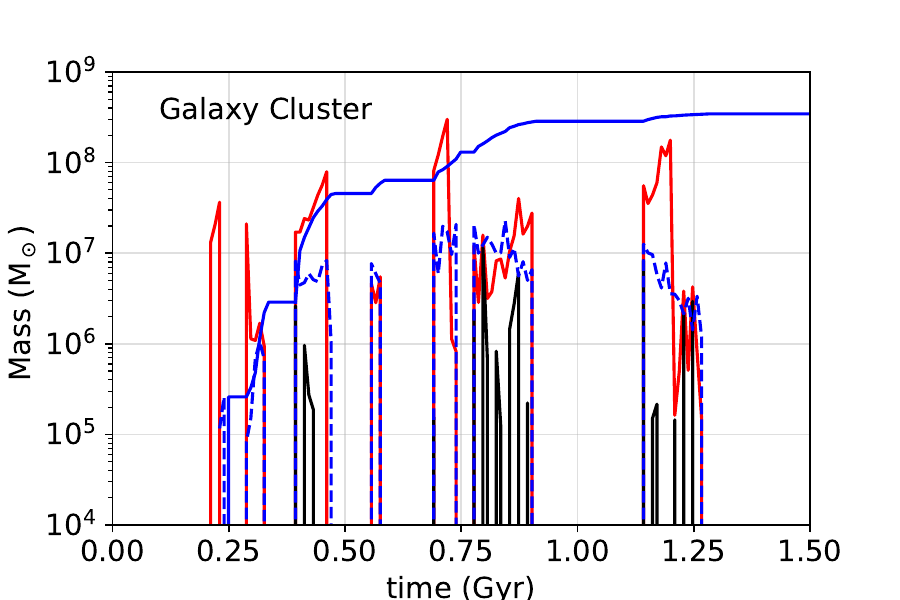}
    \caption{Cold ($T<10^5$ K) gas mass and accumulated stellar mass in the single phase galaxy (left panel), multiphase galaxy (middle panel), and brightest cluster galaxy (right panel) simulations with AGN feedback.  Red lines show the amount of cold gas, and black lines show the amount outside the central kiloparsec. Dashed blue lines shows the mass of the stars [$M_*$ (young)] formed over last 10 Myr.  Solid blue lines show the cumulative mass of stars [$M_*$ (accumulated)] formed during the course of the simulation.  No stars form in the SPG simulation.
    }
    \label{fig:temporal_evol}
\end{figure*}

In these simulations, central accumulation of cold gas couples atmospheric conditions with AGN feedback while accumulations of cold gas at larger radii facilitate star formation.  Figure \ref{fig:temporal_evol} shows how the masses of cold gas ($M_{\rm cold}$) and new stars change with time.  In the single phase galaxy, the accumulations of cold gas are always small ($10^4$--$10^6 \, M_\odot$), and star formation is negligible.  Note that the feedback algorithm described in \S \ref{sec:acc} produces an AGN feedback power
\begin{equation}
    \dot{E}_{\rm AGN} = 6 \times 10^{42} \, {\rm erg\, s^{-1}}
        \left[ \frac {M_{\rm cold}(< 0.5 \, {\rm kpc})} {10^6 \, M_\odot} \right]
        \label{eq:AGN_power}
        \; \; ,
\end{equation}
given $\epsilon_{\rm AGN} = 10^{-4}$.  The fluctuations in feedback power shown for this galaxy in Figure \ref{fig:lx_pjet} are therefore consistent with the fluctuations in cold gas mass shown in Figure \ref{fig:temporal_evol}, as long as
a large proportion of that gas ends up accreting onto the central black hole.  Cold gas clouds forming through condensation are therefore consumed before they can form stars, linking the precipitation rate within 0.5~kpc directly to AGN feedback.

In the multiphase galaxy, the mass of cold gas is $\sim 10^7 \, M_\odot$ in the high-power mode and $\sim 10^5$--$10^6 \, M_\odot$ in the low-power mode.  Those amounts of cold gas are also largely consistent with the fluctuations in feedback power shown in Figure \ref{fig:lx_pjet}.  Furthermore, the periods when cold gas extends beyond 1~kpc correlate with periods of greater star formation and AGN power.  Feedback events that cause multiphase precipitation at larger radii therefore promote star formation in our simulations, because the cold gas clouds have time to form stars before sinking into the AGN accretion region in the central 0.5~kpc. Figure~\ref{fig:nj} shows that those star-formation events briefly peak at a rate $\sim 1 \, M_\odot \, {\rm yr^{-1}}$, but the accumulated stellar mass in figure \ref{fig:temporal_evol} implies a time-averaged rate $\sim 1 \, M_\odot \, {\rm yr^{-1}}$ and a specific star-formation rate $\sim 10^{-12} \, {\rm yr}^{-1}$.

The brightest cluster galaxy experiences several condensation events that push $M_{\rm cold}$ up to $\sim 10^8 \, M_\odot$, producing surges of AGN power exceeding $10^{44} \, {\rm erg \, s^{-1}}$.  Figure \ref{fig:lx_pjet} shows that surges of this magnitude with a significant duty cycle are necessary to compensate for the radiative losses from within 30~kiloparsecs. 
Note that increasing the parameter $\epsilon_{\rm AGN}$ and accretion time in our feedback algorithm would result in greater accumulation of cold gas, because balance between AGN feedback and radiative cooling would require less cold gas to be reprocessed within the accretion zone.  
The amount of star formation in this simulation is therefore contingent on the choice of $\epsilon_{\rm AGN}$.  For the choice $\epsilon_{\rm AGN} = 10^{-4}$, the time-averaged star formation rate is $\sim 0.4 \, M_\odot \, {\rm yr}^{-1}$. 

\begin{figure}
    \includegraphics[width=0.48\textwidth]{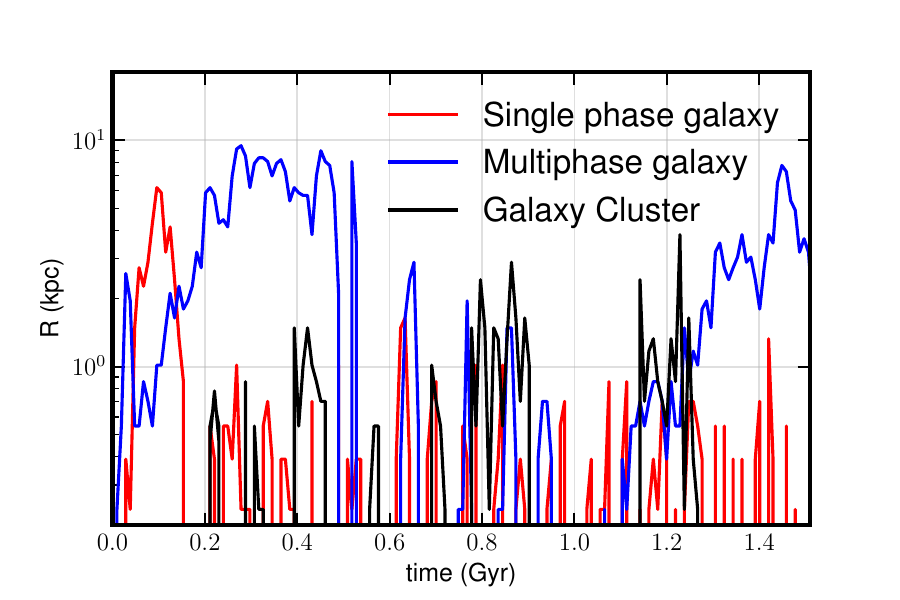}
    \caption{ Radial extent of cold gas ($T<10^5$ K) in the AGN-feedback simulations of the single phase galaxy (red line), multiphase galaxy (blue line) and brightest cluster galaxy (black line).  In the single phase galaxy, cold gas almost always remains within 1~kpc.  In all cases, cold gas reaches its maximum extent following high-power bursts of AGN feedback.}
    \label{fig:rad_ext}
\end{figure}

Figure \ref{fig:rad_ext} shows the maximum radial extent of cold gas ($T<10^5$ K) as these different systems evolve.  In the single phase galaxy simulation, cold gas remains concentrated within 1~kpc and usually within 0.5~kpc, except during the initial outburst.  In contrast, cold gas in the multiphase galaxy generally extends beyond 1~kpc (sometimes as far as $\sim 10$~kpc), but cold gas in the BCG simulation tends to be less extended.  Comparing cold gas radial extent to $P_{\rm jet}$ in Figure \ref{fig:lx_pjet} shows that cold gas becomes most extended following periods of strong AGN feedback, indicating that uplift of central gas promotes condensation at greater altitudes (\citealt{revaz2008,McN2016,voit2017}). 

\subsection{AGN feedback in a smaller halo}
\label{sec:small_halo}

\begin{figure}
    \centering
    \includegraphics[width=0.5\textwidth]{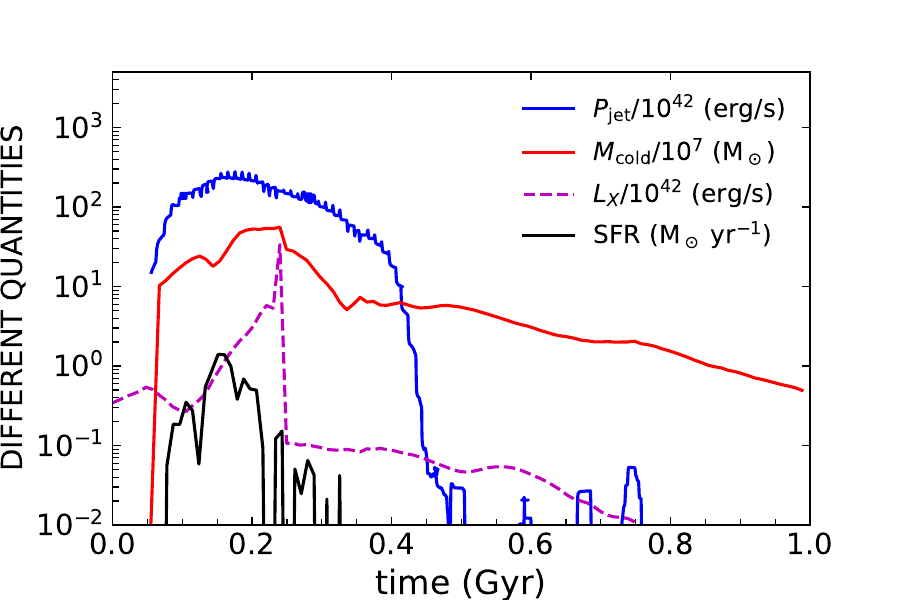}
    \caption{Temporal evolution of $P_{\rm jet}$ (red line), $M_{\rm cold}$ with 50~kpc (blue line), $L_{\rm X}$ within 50~kpc (dashed purple line), and star formation rate (black line) in the SEG simulation with AGN feedback. It does not self-regulate.  Instead, a large feedback outburst reconfigures the entire hot-gas atmosphere, leading to an extended period of gradual decline in $P_{\rm jet}$, $M_{\rm cold}$, and $L_{\rm X}$. 
    }
    \label{fig:gal_evol}
\end{figure}

Our AGN-feedback simulation in a lower-mass halo (the SEG with $M_{200} = 2 \times 10^{12} M_\odot$) dramatically differs from the others. Figure \ref{fig:gal_evol} shows that the simulation produces a large AGN outburst and some star formation during the first 0.4~Gyr and then enters a state in which AGN feedback does not compensate for radiative losses.  Instead, star formation and AGN power shut down, while $L_{\rm X}$ and $M_{\rm cold}$ steadily decline with time.

As in the SEG simulation without AGN feedback (see \S \ref{sec:nojet}), the action begins at $\sim 50$~Myr when the central gas starts to condense.  Those first cold clouds then trigger a self-exciting AGN feedback outburst.  Uplift of low-entropy ambient gas simulates multiphase condensation that causes $\sim 5 \times 10^8 \, M_\odot$ of cold gas to precipitate by $t = 200$~Myr.  Much of that cold gas falls radially back into the accretion zone ($< 0.5$~kpc), boosting the jet power by a few times -- up to $\sim 10^{44} \, {\rm erg \, s^{-1}}$, more than an order of magnitude greater than the radiative losses from the ambient medium ($L_{\rm X}$).  This powerful feedback event blows out much of the hot-gas atmosphere but does not destroy the cold gas clouds, which can continue to rain back down in the accretion zone.  The result is a decaying AGN feedback mode, in which intermittent accretion events produce smaller feedback outbursts that gradually lower the X-ray luminosity.

\begin{figure*}
    \includegraphics[width=2.35in,height=1.6in]{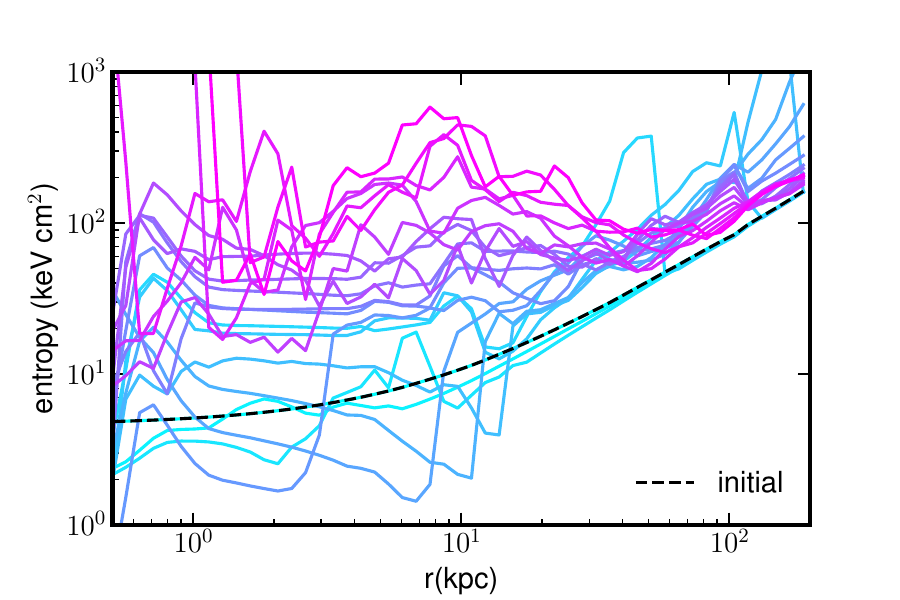}
    \includegraphics[width=2.35in,height=1.6in]{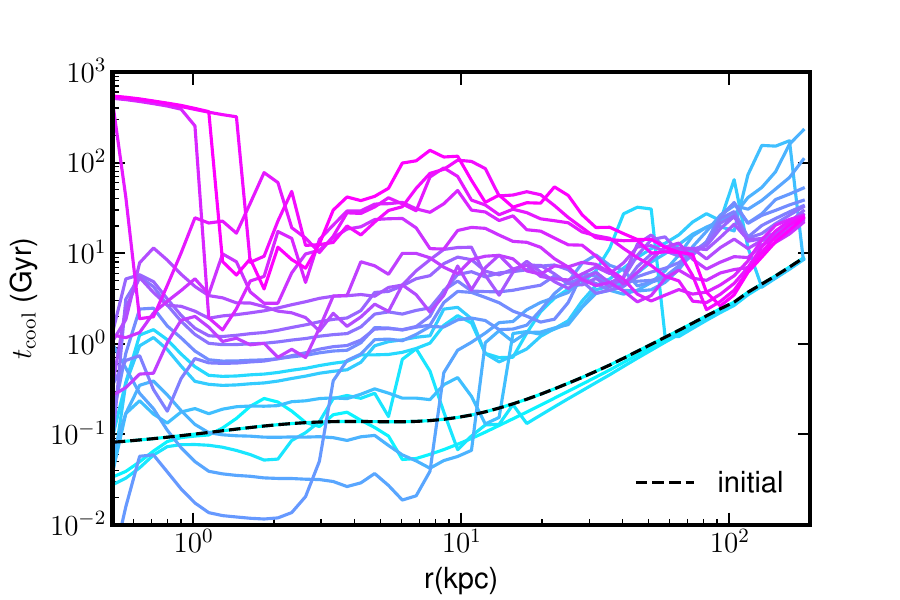}
    \includegraphics[width=2.8in,height=1.6in]{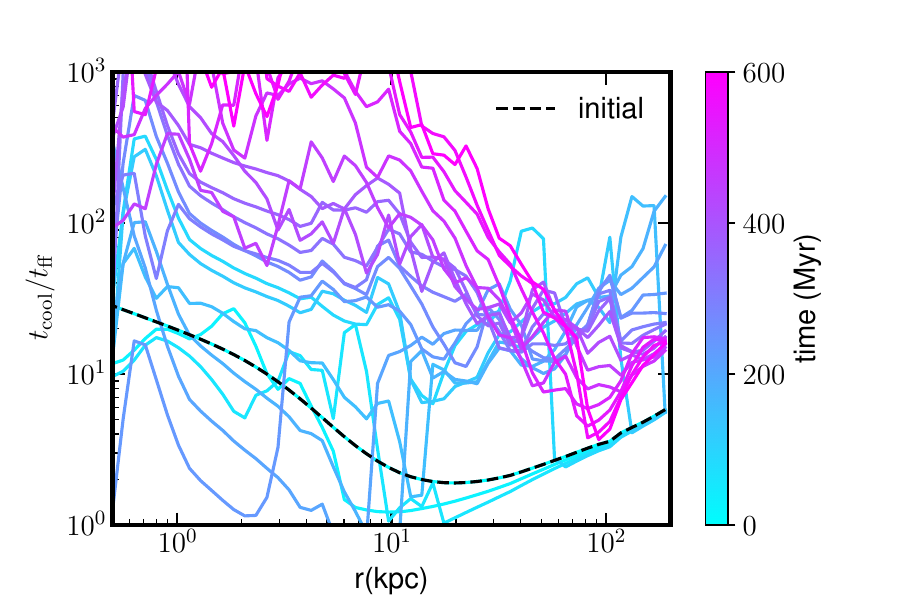}
    \caption{Evolution of the SEG simulation ($M_{200} = 2 \times 10^{12} M_\odot$) with AGN feedback in entropy (left panel), cooling time (middle panel) and $t_{\rm cool}/t_{\rm ff}$ ratio (right panel) during the first 600~Myr. Colored lines show the median radial profiles at 30 Myr
    intervals.  Central cooling initiates feedback at $\sim 50$~Myr. The AGN outflow then lifts low-entropy gas out of the central region and creates a thermally unstable entropy inversion at $\sim 10$~kpc.  Condensing low-entropy gas subsequently sinks to the center and boosts the AGN fedback power.  The resulting runaway overheats the CGM, raising its entropy level to $\sim 10^2 \, {\rm keV \, cm^2}$ out to $\sim 100$~kpc by $\sim 400$~Myr.
    } 
    \label{fig:gal}
\end{figure*}

Figure \ref{fig:gal} shows the evolution of entropy (left panel), $t_{\rm cool}$ (middle panel), and $t_{\rm cool}/t_{\rm ff}$ ratio (right panel) during the first 600 Myr of the SEG simulation with AGN feedback.  During the first 300~Myr, strong AGN feedback pushes some of the low-entropy ambient gas from the central region to beyond $\sim 10$~kpc and heats the rest.  The radiatively cooling outflow becomes most unstable to condensation near 10~kpc, where $t_{\rm cool}/t_{\rm ff}$ remains near unity for tens of Myr.  That low-entropy gas then sinks inward and becomes the cold gas that sustains runaway feedback.  A similar runaway does not happen in the more massive halos, because AGN feedback in those systems does not produce as much uplift and convection.  Later in the simulation, we find $t_{\rm cool}/t_{\rm ff} < 10$ at $\sim 100$~kpc, but that gas condenses slowly because the cooling time there is several Gyr.

\section{Limitations of our simulations}
\label{sec:limit}

The primary limitation of our AGN-feedback simulations is the thickness of the jets within the central few kiloparsecs, which is inherently a numerical limitation.  In order to model a high-velocity AGN jet in a way that is numerically stable, we need to represent the outflow by modifying a ``disk'' of cells that is several cell widths in radius.  At 1~kpc, the injected jets in our calculations subtend 1~radian, which is several times greater than the observed widths of powerful jets among the galaxies we are trying to model.  For example, the angular width of the jet in NGC~4261 is $\sim 0.2$~radian at $\sim 1$~kpc \citep{Nakahara2018}. If our simulations had comparably narrow jets with the same kinetic power they would likely drill more effectively through the gas at $\sim 1$~kpc, coupling less strongly with the local ambient medium and thermalizing less of their kinetic energy there.  We therefore hypothesize that the excess entropy at $\sim 1$~kpc in the SPG and MPG simulations, relative to the data in Figure~\ref{fig:entropy}, results from simulated jets that are too wide at that radius.  We are currently testing that hypothesis with simulations that have narrower jets.  

Another limitation of our simulations is the initial lack of angular momentum in the galactic atmosphere.  Injection of feedback energy produces turbulence that gives the cold clouds forming in that atmosphere some stochastic angular momentum, but often not enough angular momentum to prevent the clouds from sinking nearly radially down into the accretion zone at $r < 0.5$~kpc.  In the three more massive halos (SPG, MPG, and BCG), the AGN feedback mechanism nevertheless manages to self-regulate, but in the SEG simulation it does not.  We hypothesize that the initial lack of angular momentum is one of the factors that stymies self-regulation of the SEG, because it allows a self-exciting runaway of AGN feedback. 

As mentioned in \S \ref{sec:small_halo}, uplift of ambient gas by the initial AGN feedback outburst in the SEG simulation stimulates condensation of $> 10^8 \, M_\odot$ of cold gas, much of which falls directly back into the accretion zone.  Jet power therefore spikes to several times $10^{44} \, {\rm erg \, s^{-1}}$ (see Equation \ref{eq:AGN_power}), dramatically heating and disrupting the CGM (see Figure \ref{fig:gal_evol}).  However, fewer of the cold clouds condensing out of the ambient medium would fall directly into the accretion zone if the atmosphere as a whole had greater net angular momentum. More of the condensing cold gas would then settle down in a torus around the central SMBH and get decoupled from the feedback cycle \citep{Prasad15}. We are therefore preparing simulations to explore the role of angular momentum in moderating this AGN feedback mechanism in lower-mass galaxies. 

\section{Discussion}
\label{sec:disc}

\subsection{Comparison With The Feedback-Valve Model}
\label{sec:comp}

The SPG and MPG simulations presented here were designed, in part, to test the ``black-hole feedback valve" mechanism proposed by \citet{voit2015,voit2020} and summarized in \S \ref{sec:valve}.  Qualitatively, the SPG and MPG simulations with AGN feedback do indeed self-regulate as envisioned, with AGN feedback tuning itself so that local radiative cooling is similar to SNIa heating out to distances several kiloparsecs from the galaxy's center.  Figure \ref{fig:lx_pjet} shows that the SPG simulation ($\sigma_v \approx 280 \, {\rm km \, s^{-1}}$) begins with cooling exceeding heating everywhere and settles into a steady state with SNIa heating exceeding radiative cooling from $\sim 1$~kpc to $\sim 5$~kpc, as predicted by the feedback-valve model for galaxies with $\sigma_v > 240 \, {\rm km \, s^{-1}}$.   AGN feedback in this mode is fueled by cooling of gas within the central 0.5~kpc, while SNIa heating sweeps much of the gas released by stars at larger radii out of the galaxy.  This state can remain steady as long as AGN feedback prevents the confining CGM pressure from building up, and it succeeds for at least 1.5~Gyr because the time-averaged jet power roughly matches radiative losses from the inner 30~kpc. However, we have not yet tested whether a galaxy with the SPG potential but a higher-pressure CGM (like the initial state of the MPG) tunes itself to this same steady state.

The MPG simulation ($\sigma_v \approx 230 \, {\rm km \, s^{-1}}$) also begins with cooling exceeding SNIa heating everywhere.  Kinetic AGN feedback with power $\sim 10^{44} \, {\rm erg \, s^{-1}}$ then lowers the atmospheric density and abates when radiative cooling becomes similar to SNIa heating within $\sim 3$ kpc.  The galaxy remains in this low-power state for nearly 1~Gyr, but cannot sustain it because radiative losses from the inner 30~kpc exceed the jet power.  The CGM pressure there remains high and gradually increases, preventing supernova heating from sweeping ejected stellar gas out of the galaxy.  Gas density and radiative cooling within the galaxy therefore both rise until the gas density reaches a ceiling imposed by the condition $\min(t_{\rm cool} / t_{\rm ff}) \approx 10$ (see Figure~\ref{fig:tcool}), at which precipitation of cold clouds inevitably triggers a large increase in feedback power.  This second burst of feedback again lowers the central gas density until SNIa heating is comparable to local radiative cooling (see the yellow line in the lower center panel of Figure~\ref{fig:lx_pjet}).  Consequently, the configuration of the ambient medium fluctuates but is bracketed by the state in which SNIa heating exceeds radiative cooling and the state with $\min (t_{\rm cool} / t_{\rm ff} ) \approx 10$, in alignment with the black-hole feedback valve model for galaxies with $\sigma_v \approx 230 \, {\rm km \, s^{-1}}$.

Quantitatively, however, the simulations do not match the black-hole feedback valve model in detail.  One of the model's key predictions is that the power-law slope of the entropy profile should exceed $K \propto r^{2/3}$ at $r \sim 1$--10~kpc in galaxies with $\sigma_v > 240 \, {\rm km \, s^{-1}}$.  The entropy profile at 1--10~kpc in our SPG simulation is much flatter than this prediction, which assumes that SNIa heating exceeds AGN heating within the galaxy.  While it is difficult to measure with precision how much of the heating at $\sim 1$~kpc is resulting from thermalization of jet energy, the excess entropy at that radius in the SPG simulation, relative to both the data and the analytical steady-flow models of \citet{voit2020}, suggests that the discrepancy results from excessive thermalization of jet kinetic energy at $\sim 1$~kpc in the simulation (see \S \ref{sec:limit}).

The predictions made in \citet{voit2020} for galaxies like the SEG and BCG are less specific, but the results of our simulations of the SEG and BCG with AGN feedback generally conform to the model's expectations.  According to the model, a galaxy that does not supply enough feedback power to lift its CGM and alleviate the confining pressure should remain in a precipitation-limited state that self-regulates through multiphase condensation.  The BCG simulation is consistent with that expectation of the model.  For the SEG, the model predicts that multiphase circulation should be inevitable, because feedback overturns the atmosphere's entropy gradient.  And indeed, the feedback events observed in the simulation lift low-entropy ambient gas out of the center, catalyzing widespread condensation and production of cold gas, much of which falls back toward the center.

\subsection{Comparisons with Prior Simulations}
\label{sec:PreviousSims}

Several earlier numerical studies have explored the role of kinetic AGN feedback fueled by cold-gas accretion in massive elliptical galaxies like the ones simulated in this paper.  The efforts most closely related to our SPG and MPG simulations were published by \citet{gaspari2011,Gaspari2012_ellipticals} and \citet{wang2019}.  The ones most closely related to our BCG simulation were published by \citet{Gaspari2011_clusters,gaspari2012},\citet{li15}, \citet{Prasad2018}, and \citet{meece2017}.

\citet{gaspari2011,Gaspari2011_clusters,Gaspari2012_ellipticals} performed the first suite of simulations to demonstrate that bipolar jets fueled by cold accretion can tune themselves to balance radiative cooling without overheating the central gas.  Collectively, those three papers explored a range of halo mass similar to the range spanned by our SPG, MPG, and BCG models, but with substantially lower spatial resolution. Also, they adjusted their AGN feedback efficiency parameter, equivalent to our $\epsilon_{\rm AGN}$, to optimize agreement with observations, finding the best results for $\epsilon_{\rm AGN} \lesssim 3 \times 10^{-4}$ in lower-mass halos and $\epsilon_{\rm AGN} \gtrsim 5 \times 10^{-3}$ in cluster-scale halos.  While mass and energy input from the old stellar population were included in these simulations, the role of the old stellar population in the overall feedback loop was not specifically analyzed.

The simulations of \citet{wang2019}, like ours, were motivated by the analysis of \citet{voit2015} and focused on distinguishing the roles of the central gravitational potential and the stellar mass and energy sources.  \citet{wang2019} performed two simulations similar to our SPG and MPG simulations with AGN feedback.  The initial conditions in those simulations were not identical to ours but were inspired by the same two galaxies, with NGC 4472 representing single phase elliptical galaxies and NGC 5044 representing multiphase galaxies.  In alignment with our simulation results, \citet{wang2019} found that AGN feedback in the galaxy similar to NGC~4472 maintained a relatively steady hot-gas atmosphere with small amounts of centrally concentrated cold gas, while the same AGN feedback algorithm in the galaxy similar to NGC~5044 caused greater fluctuations in the hot-gas atmosphere and produced larger quantities of extended cold gas.  Their general findings therefore also support the black-hole feedback valve model.

However, the details of our simulation results differ from those of \citet{wang2019}. First, the median AGN power in \citet{wang2019} is much greater, with jet power often rising above $10^{43} \, {\rm erg \, s^{-1}}$ in the single phase galaxy and rarely dropping below $10^{43} \, {\rm erg \, s^{-1}}$ in the multiphase galaxy.  Their large AGN feedback efficiency (equivalent to $\epsilon_{\rm AGN} = 5 \times 10^{-3}$) allows the same amount of cold-gas accretion to produce much more power, but that cannot be the whole explanation for the power difference, because self-regulation over a 1.5~Gyr period requires time-integrated heat input within the central 10 to 30~kpc (with $t_{\rm cool} \lesssim 1.5$~Gyr) to balance radiative losses from the same region.  Therefore, kinetic AGN power in \citet{wang2019} must be thermalizing over a larger region, implying that it is propagating farther from the center.  Second, specific entropy near $\sim 1$~kpc in the single phase galaxy of \citet{wang2019} remains below $10 \, {\rm keV \, cm^2}$ most of the time, and is typically $\approx 5 \, {\rm keV \, cm^2}$, in better agreement with observations of single phase galaxies than our SPG simulation.  In \S \ref{sec:limit}, we hypothesized that the excess entropy at $\sim 1$~kpc in our SPG simulation resulted from jets that were insufficiently narrow.  And indeed, the jets implemented by \citet{wang2019} are narrower, having a transverse momentum profile $\propto \exp (-r^2 /2 r_{\rm jet}^2)$ with $r_{\rm jet} = 183$~pc.  With greater power and a smaller cross-section, the jets in \citet{wang2019} have a much greater momentum flux than ours and are capable of propagating to much greater distances, also accounting for why the simulations of \citet{wang2019} require more AGN power to self-regulate.

Self-regulation of AGN feedback in our BCG simulation is broadly similar to what is observed in other simulations of its type \citep[e.g.,][]{Gaspari2011_clusters,gaspari2012,li2014,li15,Prasad15,Prasad2018,meece2017}.  Our BCG simulation's typical value of $\min (t_{\rm cool} / t_{\rm ff})$ is greater than most, with a median ratio $\sim 25$.  In a future paper, we will show that the self-regulated $K(r)$ profile of our BCG simulation, which has an inner slope $K \propto r^{2/3}$ (see Figure~\ref{fig:entropy}) is in excellent agreement with the observations of \citet{Hogan2017} and \citet{babyk2018}.  However, unlike some of the other simulations, it produces less cold gas than is observed in cluster cores, and the cold gas it does produce rarely extends beyond 3~kpc.  

The main reason for the lack of cold gas in our BCG simulation is the low feedback efficiency parameter and the small accretion time we have chosen.
The small accretion time causes the cold gas forming within $r<0.5$ kpc to be quickly removed from the simulation domain. This results in a high accretion rate, producing powerful AGN jets before much cold gas accumulates in the cluster core, despite the low AGN feedback efficiency. For $\epsilon_{\rm AGN} = 10^{-4}$, the cold-gas accretion rate required to sustain $10^{44} \, {\rm erg \, s^{-1}}$ of feedback power is $18 \, M_\odot \, {\rm yr^{-1}}$.  Our algorithm converts all of that cold gas to hot gas and expels it from the central region in a jet.  Over the course of 1~Gyr, more than $10^{10} \, M_\odot$ would otherwise have accumulated, and much it would likely have formed stars at a rate $\sim 10 \, \mathrm{M}_\odot \, \mathrm{yr}^{-1}$. In some of the other cluster-scale simulations \citep[e.g.,][]{gaspari2012,li2014,Prasad15}, much of the cold gas persists indefinitely in a torus orbiting outside of the accretion zone because of the stochastic angular momentum it gains during kinetic feedback bursts.  Our simulation does not produce such a torus and we will analyze what inhibits torus formation in a future paper.

\section{Conclusions}
\label{sec:conc}

The suite of simulations in this paper was designed to explore how a particular cold-fueled kinetic AGN feedback mechanism responds to differences in the surrounding potential well and initial atmospheric conditions.  In halos ranging from galaxy-cluster scale ($8\times10^{14} M_\odot$), through galaxy-group scale ($4\times10^{13} M_{\odot}$), down to smaller elliptical galaxies ($2\times10^{12} M_\odot$), we performed high resolution 3D hydrodynamic simulations with radiative cooling, stellar feedback, and AGN feedback.  We were particularly interested in testing the ``black-hole feedback valve" mechanism (see \S \ref{sec:valve}), which hypothesizes that coupling between AGN feedback and SNIa heating tunes the confining CGM pressure so that SNIa heating approximately equals radiative cooling within the galaxy. 

The main results from those numerical experiments are:
\begin{enumerate}

    \item AGN feedback is necessary to quench star formation in all of our simulated galaxies.

    \item The cold-fueled kinetic AGN feedback mechanism we implement becomes self-regulating within $\sim 200$~Myr in all three of the higher mass halos ($M_{200} > 10^{13} \, M_\odot$).
    
    \item AGN feedback in the two group-scale halos self-tunes to a state with SNIa heating approximately equal to radiative cooling inside the central galaxy, and the nature of that self-regulated state depends on galactic velocity dispersion ($\sigma_v$) and confining CGM pressure. Those findings, which mirror those of \citet{wang2019}, are in general agreement with the black-hole feedback valve hypothesis.
    
    \item AGN feedback in our single phase galaxy (SPG) simulation with $\sigma_v \approx 280$ km s$^{-1}$ maintains a nearly steady state, with time-averaged AGN power several times $10^{41} \, {\rm erg \, s^{-1}}$.  Condensation of cold gas is focused within the central kiloparsec, as predicted by the black-hole feedback valve model for galaxies with $\sigma_v > 240 \, {\rm km \, s^{-1}}$.  SNIa heating exceeds radiative cooling at $\sim 1$--5~kpc and sweeps much of the gas ejected by stars out of the galaxy, while star formation is completely quenched.  However, kinetic AGN feedback appears to overheat the region near $\sim 1$~kpc, producing excess entropy, relative to observations.  We hypothesize that the bipolar jets implemented in our simulations overheat that region because they are too wide and therefore couple too strongly to the ambient gas there.
    
    \item AGN feedback in our multiphase galaxy (MPG) simulation with $\sigma_v \approx 230$ km s$^{-1}$ is less steady, switching back and forth between a high-power state ($\sim 10^{44} \, {\rm erg \, s^{-1}}$) and a low-power state ($\sim 10^{42} \, {\rm erg \, s^{-1}}$).  The high-power state is characterized by $\min (t_{\rm cool} / t_{\rm ff} ) \sim 10$, extending to $\sim 15$~kpc, which allows precipitation of cold clouds out of the hot ambient medium to produce extended multiphase gas. AGN power fueled by accretion of the cold gas then heats the CGM and lowers its pressure until SNIa heating approximately matches radiative cooling within the central few kiloparsecs.  As that happens, the MPG simulation enters a low-power state but cannot maintain it, eventually reverting back to the high-power state with $\min (t_{\rm cool} / t_{\rm ff} ) \sim 10$.  These features are consistent with the black-hole feedback valve model for galaxies with $\sigma_v \lesssim 240 \, {\rm km \, s^{-1}}$.
    
    \item CGM pressure in our brightest cluster galaxy (BCG) simulation is always great enough to ensure that radiative cooling exceeds SNIa heating everywhere.  It self-regulates with AGN power exceeding $10^{44} \, {\rm erg \, s^{-1}}$ for much of the simulation runtime.  However, not much cold gas accumulates compared to other similar galaxy-cluster simulations, probably because of our comparatively low feedback efficiency parameter ($\epsilon_{\rm AGN} = 10^{-4}$).  
    
    \item In the smaller elliptical galaxy (SEG) simulation, with $\sigma_v \approx 150 \, {\rm km \, s^{-1}}$, the cold-fueled kinetic feedback mechanism dramatically fails to self-regulate. As AGN feedback turns on and begins to lift the ambient gas, it stimulates copious multiphase condensation.  Much of that cold gas then rains back down into the accretion region, causing an even stronger feedback response.  This runaway of AGN feedback then overheats the ambient gas and blows out much of it out to $\sim 100$~kpc. We suspect that the outcome of this simulation might have been different if the galaxy's initial atmosphere had some net angular momentum.  Much of the precipitating cold gas might then have avoided falling into the accretion zone and fueling the runaway response.

\end{enumerate}

\noindent Future papers will present more detailed analyses of each of these simulations.

\acknowledgments
DP is supported by {\it Chandra} theory grant no. TM8-19006X (G. M. Voit as PI) and NSF grant no. AST-1517908 (B.W.O'Shea as PI). BWO acknowledges further support from NSF grant no. AST-1908109 and NASA ATP grants NNX15AP39G and 80NSSC18K1105.
DP thanks Yuan Li and Philipp Grete for providing support during the start of this project.
This work used the Extreme Science and Engineering Discovery Environment (XSEDE), which is supported by National Science Foundation grant number ACI-1548562 and TG-AST190022, as well as the resources of the Michigan State University High Performance Computing Center (operated by the Institute for Cyber-Enabled Research). 
\texttt{Enzo} \citep{Bryan2014,Enzo_2019} and
\texttt{yt\textbf{}} \citep{YT} are developed by a large number of independent researchers from
numerous institutions around the world. Their commitment to open science
has helped make this work possible.

\bibliographystyle{aasjournal}
\bibliography{reference}
\newpage

\begin{appendix}

Section \S \ref{sec:limit} of the paper discusses one of the key limitations of our simulations.  The AGN jets we have implemented have a wide opening angle ($\theta \approx 1$ radian), which generally results in thermalization of jet kinetic energy at smaller radii compared to narrower jets.  That is a plausible reason for the excess entropy at $r<2$ kpc for our single phase galaxy simulations, when compared to observations of single phase elliptical galaxies (see the left panel of Figure \ref{fig:entropy}). 

\begin{figure*}[h]
    \includegraphics[width=2.35in,height=1.6in]{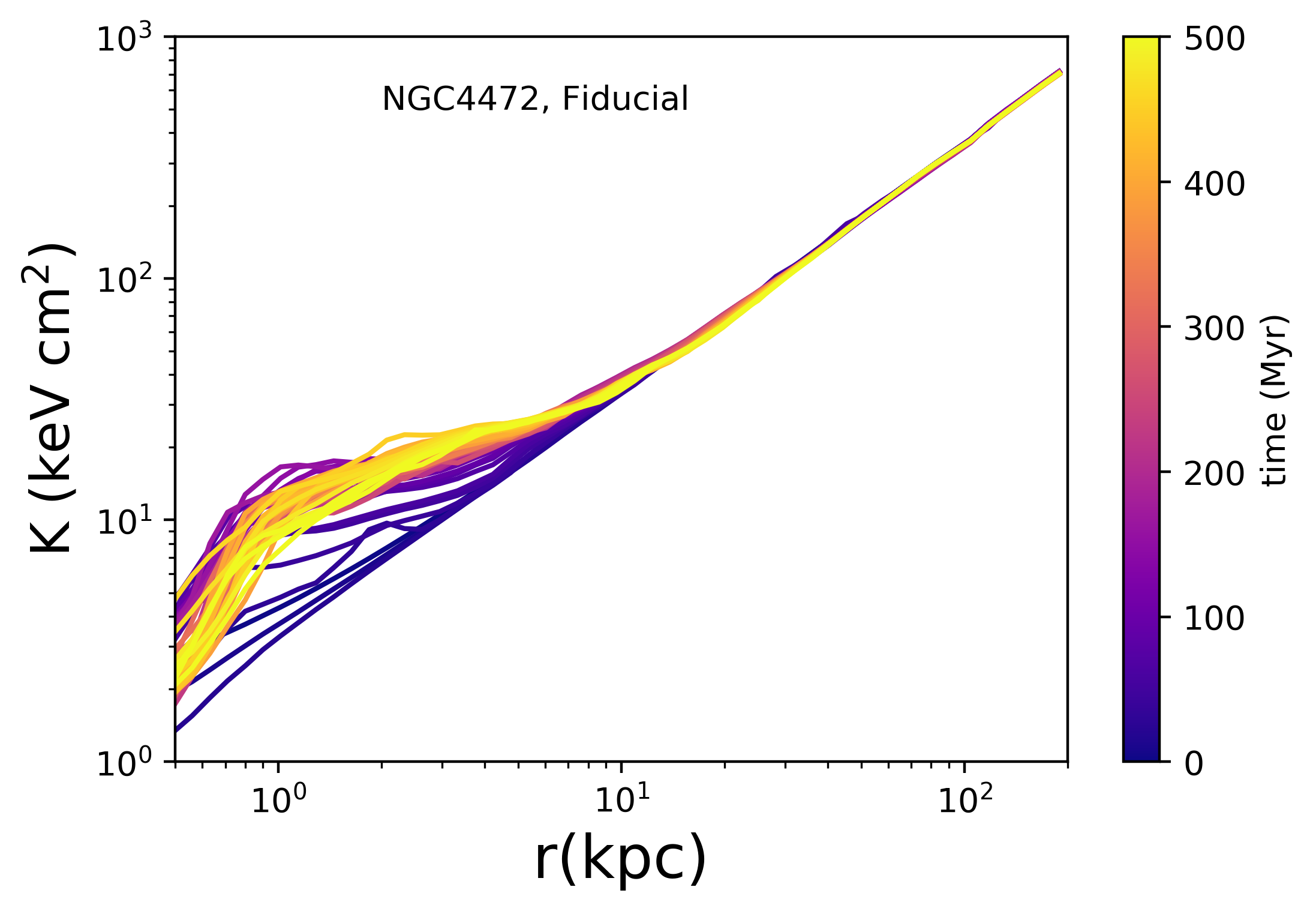}
    \includegraphics[width=2.35in,height=1.6in]{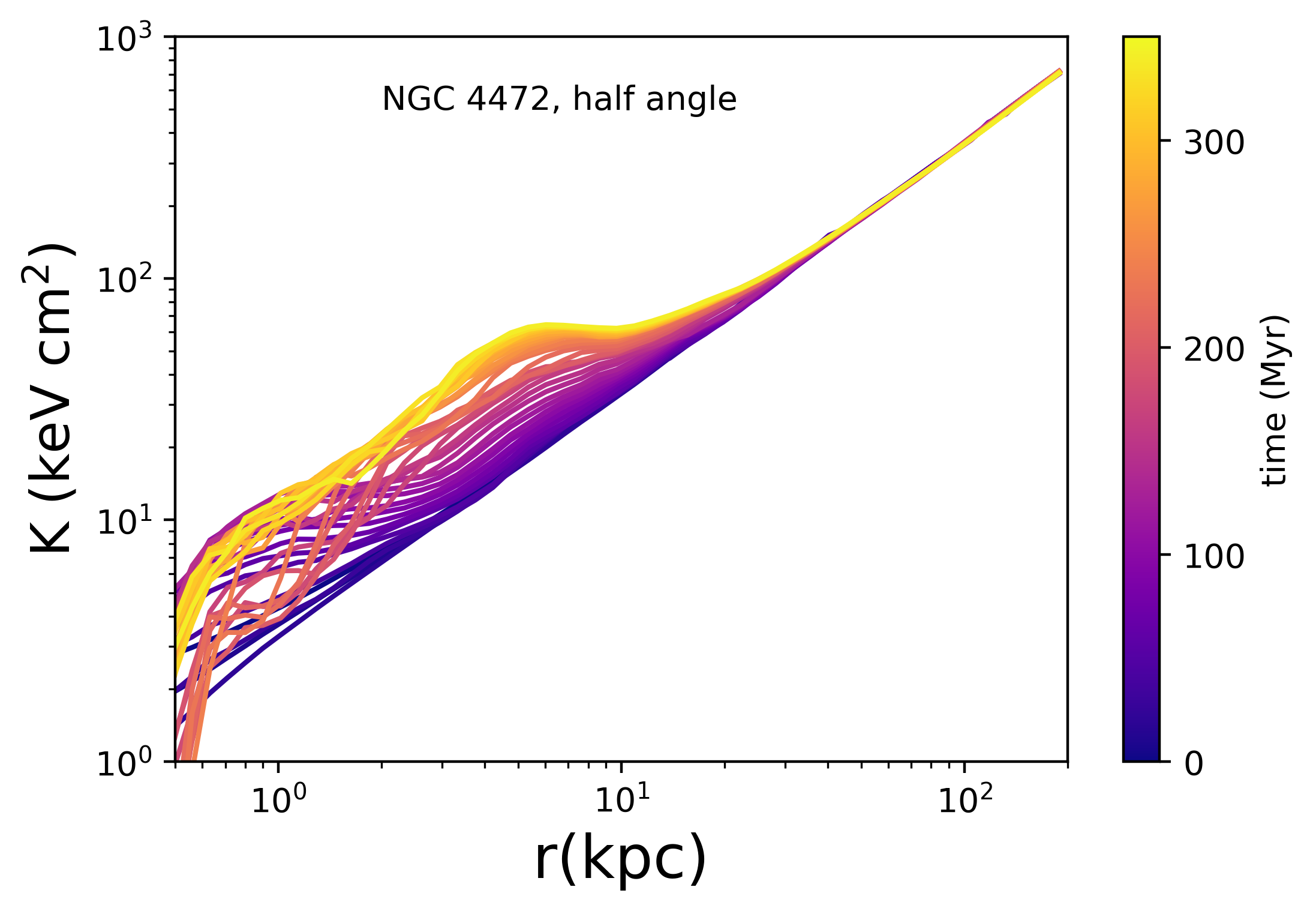}
    \includegraphics[width=2.35in,height=1.6in]{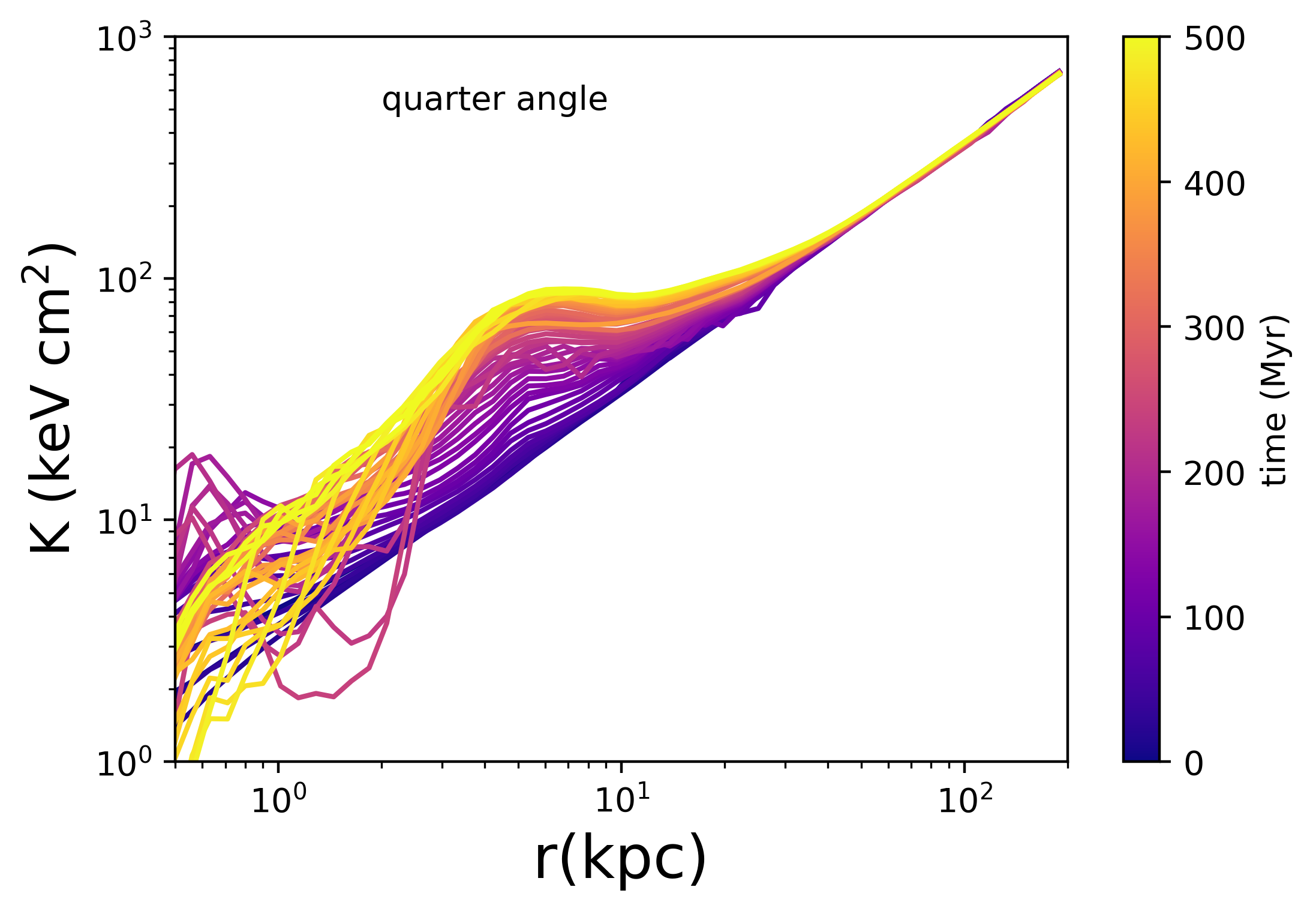}
    \caption{Effects of changing jet width on entropy profile evolution.  All three panels show simulations of the single phase galaxy resembling NGC~4472.  The left panel shows our fiducial simulation, in which the jets have an opening angle of 1~radian at the injection radius.  In the center panel, the jets have an opening angle of 0.5~radians.  In the right panel, the jet opening angle is 0.25~radians.  The greater increases in entropy at large radii in the center and right panels demonstrate that narrower jets propagate farther and thermalize their kinetic energy at greater radii (10--30 kpc).  As a result, entropy levels at smaller radii (1--3 kpc) generally remains lower, because less of the jet energy thermalizes there.
    } 
    \label{fig:n4472}
\end{figure*}

After submission of the original manuscript, we initiated a systematic exploration of the effects of jet width on our simulated entropy profiles.  Figure \ref{fig:n4472} shows some preliminary results.  Its left panel shows our fiducial SPG simulation, with a jet opening angle of 1~radian at the jet injection radius (1~kpc).  The center panel shows what happens when that opening angle is cut in half, to 0.5~radian.  An entropy increase is seen larger radii (10--20 kiloparsecs), confirming that the jets propagate farther and thermalize more of their kinetic energy at greater radii.  The right panel shows what happens when the opening angle is further reduced to 0.25~radian.  Significant entropy increases are then seen at even greater radii ($\sim 30$~kpc).  Also, in both simulations with narrower jets, entropy levels at 1--3 kiloparsecs are generally smaller than in the fiduical simulations and are in better agreement with the observed entropy profiles of single-phase galaxies.

We will present a more detailed analysis of the effects of changing jet width in a forthcoming paper that focuses exclusively on SPG and MPG simulations and explores their characteristics in greater detail.

\end{appendix}

\end{document}